\documentclass[journal]{IEEEtran}
\IEEEoverridecommandlockouts
\pdfoutput=1
\usepackage{amsmath,amssymb,amsfonts}
\usepackage{mathtools}
\usepackage{algorithm}
\usepackage{algorithmic}
\usepackage{textcomp}
\usepackage{xcolor}
\usepackage{array}
\usepackage{stfloats}
\usepackage{url}
\usepackage{float}
\usepackage{graphicx}
\usepackage{verbatim}
\usepackage{cite}
\usepackage{booktabs}
\usepackage{enumerate}
\usepackage{boondox-cal}
\usepackage{bm}

\DeclareMathOperator*{\argmin}{argmin}

\def\BibTeX{{\rm B\kern-.05em{\sc i\kern-.025em b}\kern-.08em
    T\kern-.1667em\lower.7ex\hbox{E}\kern-.125emX}}
\begin{document}

\title{Robust Resource Management for SAGIN using DNN-Driven Channel Uncertainty Learning}

\author{Haonan Zhang, Weihua Wu, Qi Zhang, Runzi Liu, Zewei Jing, Weijia Han
\thanks{This work was supported in part by the Natural Science Basis Research Planin Shaanxi Province of China (Grant No. 2025JC-YBMS-674), in part by the TechnologicalInnovation Guidance Program (Fund)- Three Reforms - Performance-Based Evaluation (Grant No. 2025ZC-YYDP-51), in part by the Shaanxi Key Industrial Innovation ChinaProject in Industrial Domain (Grants No.2023-ZDLGY-15, No.2023-ZDLGY-51), in part by ``Scientist + Engineer" team building in Qinchuangyuan (Grant No. 2025QCY-KXJ-169)and in part by the Key Research and Development Program of Shaanxi (Program No.2025CY-YBXM-059).}
\thanks{Haonan Zhang, Weihua Wu, and Weijia Han are with the School of Physics and Information Technology, Shaanxi Normal University, Xi'an 710119, China (emails: zhanghaonan@snnu.edu.cn, whwu@snnu.edu.cn, wjhan@snnu.edu.cn).}
\thanks{Qi Zhang is with the School of Telecommunication Engineering, Xidian University, Xi'an 710126, China(dakexiaoqi@stu.xidian.edu.cn).}
\thanks{Runzi Liu is with the School of Telecommunications Engineering, Xi'an University of Architecture and Technology, Xi'an 710055, China (email: rz\_liu@126.com).}
\thanks{Zewei Jing is is with the Guangzhou Institute Technology, Xidian University, Guangzhou 510555, China, and also with the State Key Laboratory of Internet of Things for Smart City, University of Macau, Macau SAR, China (email:
jingzewei@xidian.edu.cn).}
}

\maketitle
\begin{abstract}
This paper focuses on the joint robust beamforming and resource allocation for space-air-ground integrated networks (SAGIN) under uncertain channel state information (CSI). In SAGIN, uncertain CSI undermines the precise adjustment of beamforming and resource allocation, posing a major challenge to meeting heterogeneous users' strict quality of service (QoS) requirements. 
To address this challenge, we first formulate a chance-constrained optimization problem to minimize the total transmit power while satisfying QoS requirements under a predefined outage probability. By leveraging semidefinite relaxation (SDR), the objective function is transformed into a linear function of the traces of the beamforming matrices. Then, we propose a deep neural network (DNN)-driven channel uncertainty learning to dynamically learn and model the uncertain CSI as an asymmetric uncertainty set. Under the constructed CSI uncertainty set, a robust counterpart method based on pre-trained network parameters is developed. It characterizes the uncertainty set as a finite union of convex sets, thereby providing a tractable approximation for the original chance constraints.
Finally, we design an adaptive iterative algorithm to jointly optimize the resource allocation and beamforming vectors. Simulation results show that our proposed DNN-driven method outperforms traditional robust and non-robust methods, achieving a superior energy efficiency and robust reliability in SAGIN.
\end{abstract}
\section{\uppercase{{\large I}ntroduction}}
With the global maturity of fifth-generation (5G) networks, sixth-generation (6G) wireless system research has entered a critical stage of technological innovation. Academic efforts have focused on core 6G architectures including distributed autonomous networks \cite{9083888}, space-air-ground integrated networks (SAGIN) \cite{10599519,9904508}, native computing \cite{9205980}, and network digital twins \cite{10283539}. Among them, SAGIN is an indispensable paradigm for 6G massive connectivity and ubiquitous coverage \cite{10745905}. 
Under the vision of SAGIN, the scale of serviced user nodes expands rapidly with extensive spatial coverage. Nevertheless, restricted by stringent physical payload and wireless resource limitations (e.g., Size, Weight, and Power (SWaP) constraints) in the SAGIN platform, conventional orthogonal multiple access (OMA) technologies fail to well balance the inherent contradiction between the demand for massive user access and limited spectral efficiency, owing to their rigid orthogonality restriction on resource allocation. To break this bottleneck, this paper adopts beamforming-driven non-orthogonal multiple access (NOMA) as a key enabling scheme \cite{9711564,8528847}. The proposed architecture exploits beamforming to implement high-gain directional transmission and inter-layer interference suppression in the spatial domain. On this basis, incorporates the power-domain multiplexing mechanism of NOMA to achieve intensive sharing of time-frequency resources within the same beam. This spatial-power multidimensional resource multiplexing paradigm effectively circumvents the quantity limitation of orthogonal resource blocks (ORBs), which not only considerably enhances the massive access capacity of the system, but also significantly improves the spectral efficiency and overall system throughput in resource-constrained environments.

The successful implementation of NOMA-based resource management and accurate spatial beam alignment for beamforming both rely heavily on high-precision channel state information (CSI). Unfortunately, obtaining perfect CSI is practically infeasible in dynamic SAGIN systems due to inherent physical and geographical constraints. For space-to-ground links, the ultra-long propagation distances inherently cause substantial round-trip times (RTT). Coupled with the high velocities of satellites, this introduces significant Doppler shifts, rendering the CSI feedback collected at the transmitter significantly outdated and mismatched with instantaneous channel states \cite{10750262}. These aggregated physical impairments considerably degrade the received signal-to-interference-plus-noise ratio (SINR) \cite{7482050,9345124} and inevitably generate substantial CSI estimation errors across the entire network.
In practice, relying on such imperfect CSI forces the system into two critical issues: channel underestimation and overestimation. If the channel quality is underestimated, the system assumes a pessimistic scenario and adopts an overly conservative transmission strategy, resulting in a substantial waste of scarce SAGIN resources. Conversely, if the channel is overestimated, the system allocates insufficient transmit power, inevitably causing frequent QoS violations, service interruptions and communication outages. To fundamentally avoid these adverse consequences, it is a critical imperative to explicitly model the statistical distribution of CSI and construct precise channel uncertainty sets, thereby mathematically bounding the estimation inaccuracies to guarantee robust transmissions.

To address the construction of CSI uncertainty sets in complex SAGIN, we utilize the basic spatial partitioning capability of deep neural networks (DNN) to encapsulate CSI uncertainty. A DNN equipped with continuous piecewise affine activation functions (e.g., Rectified Linear Unit, ReLU) can be inherently regarded as a geometric partitioner for high-dimensional spaces. Through the combination of linear transformation and nonlinear activation, each layer of the network constructs a set of interwoven hyperplanes. After multi-layer stacking, the composite effect of the activation functions enables the recursive partitioning and nonlinear folding of the input space. From the perspective of global mapping properties, the output of a DNN is equivalent to a complex piecewise linear manifold formed by the seamless tessellation of a large number of tiny convex polyhedra. Benefiting from this inherent geometric property, DNNs are capable of performing compact and flexible nonlinear modeling and characterization of the bounded uncertainty sets existing in CSI, thereby providing effective data-driven modeling capability for robust resource management.

Based on these motivations, we propose a chance-constrained robust resource management framework aided by DNN-driven channel uncertainty learning. 
We consider a heterogeneous SAGIN architecture, where a  satellite serves multiple satellite terminals (STs) in each beam via NOMA, and a low-altitude aerial platform (AP) serves ground internet of things devices (IoTDs) via layered division multiplexing (LDM). To maximize the overall spectral utilization of the network, a full frequency reuse scheme is adopted across the satellite and AP, which inevitably introduces cross-tier interference. Specifically, the downlink transmissions from the AP inflict severe interference on the satellite-served STs, while the system also faces inherent intra-beam and inter-beam interference from the satellite network. To guarantee energy-efficient and reliable transmissions, we formulate a chance-constrained optimization problem to minimize the total transmit power under heterogeneous QoS requirements.
The key contributions of this paper can be summarized as follows:
\begin{itemize}
\item A chance-constrained robust resource management framework is proposed to handle the joint resource management and beamforming with uncertain CSI in SAGIN. The proposed framework can optimally minimize the total transmit power of the whole network while strictly guaranteeing the heterogeneous QoS requirements of different users.
\item A DNN-driven uncertainty set modeling method is developed to achieve the asymmetric and tight characterization of CSI uncertainty. By utilizing continuous piecewise affine activation functions, the spatial distribution of uncertain CSI can be partitioned into a finite number of convex regions. This approach can extract high-dimensional features from historical channel data to closely fit the irregular uncertainty distribution and eliminate geometric redundancy.
\item A two-layer iterative resources optimization algorithm is proposed to achieve coordinated optimization of power allocation and beamforming vectors. The outer layer dynamically optimizes the NOMA power allocation coefficients using a coarse-to-fine precise search strategy, while the inner layer employs the cutting plane method (CPM) to solve the robust beamforming sub-problem. This framework converts the intractable semi-infinite constraints into a solvable sequence of finite constraints, ensuring reliable QoS guarantees and achieving high energy efficiency.
\end{itemize}

The rest of this paper is organized as follows. Section II reviews related works. Section III presents the system model, problem formulation and simplification. Section IV proposes the robust optimization framework based on DNN-driven uncertainty set learning and elaborates on the two-layer iterative algorithm for solving the joint resource allocation and beamforming problem. Subsequently, Section V provides the simulation results. Finally, Section VI concludes this paper.
\section{\uppercase{{\large R}elated {\large W}orks}}
Extensive research has been conducted on resource management and beamforming design in SAGIN. To address the problem of low-overhead resource management in satellite communications, various beamforming resource allocation schemes have been proposed to maximize spectral efficiency \cite{9674687,9485040}. Furthermore, to address the critical challenges of time-varying link qualities in integrated terrestrial-satellite networks, handover-enabled resource optimization and joint power-bandwidth allocation frameworks have been actively investigated \cite{9984697,10398511}. Recently, neural networks have also been leveraged to enhance network utility \cite{9866823,9316937}; for instance, a hypergraph neural network (HGNN)-enhanced reinforcement learning approach was developed in \cite{11148110} to manage dynamic interference and allocate resources in satellite systems, while \cite{10418568} integrated convex optimization with deep learning to enable joint channel and power allocation for satellite downlink networks, both achieving improved transmission rates and resource efficiency. Despite their theoretical significance and improvements in network utility, these works rely on the idealized assumption of perfect CSI. In practical space-to-ground communication scenarios, severe feedback delays and persistent channel estimation errors are inevitable. Consequently, these non-robust algorithms are highly susceptible to severe performance degradation and frequent communication outages in non-ideal environments.

To mitigate the detrimental effects of imperfect CSI, significant research efforts have been devoted to robust transmission design. Traditional robust optimization approaches characterize CSI uncertainties as predefined symmetric geometric sets, such as bounded boxes, ellipsoids, or simple polyhedra \cite{8170968,7089306}. While these models are mathematically tractable and easily integrated into standard convex solvers, their structural symmetry inherently restricts their flexibility. Distinct from traditional designs, other methods \cite{10336551,6364471,10678835} tackles CSI imperfection by considering the worst-case CSI distribution to satisfy outage probability constraints. However, both the predefined symmetric constraints and the rigid worst-case formulations tend to be overly conservative, since they inherently fail to capture the asymmetric and irregular distribution patterns of practical CSI. This structural discrepancy compels the system to provision resources for extreme scenarios that rarely occur, leading to excessively conservative strategies and severe resource waste. Alternatively, stochastic optimization approaches based on Lyapunov frameworks or stochastic differential equations (SDE) have been developed to address dynamic CSI variations \cite{9452072,8861400}. However, these techniques require instantaneous or high-frequency CSI acquisition, which is practically infeasible for long-delay SAGIN links. Moreover, approaches relying purely on statistical characteristics \cite{10409505,10918973} depend heavily on the accuracy of assumed distributions, which restricts their applicability in highly dynamic and uncertain SAGIN environments.

To break the bottlenecks of traditional robust optimization, recent studies have started exploring machine learning-based methods to construct uncertainty sets. For example, support vector clustering (SVC)-based methods have been proposed in \cite{9720998,11240144, 11342326} to characterize the CSI uncertainty in wireless networks. However, these techniques suffer from a fundamental limitation, since they rely on constructing a single, monolithic convex hull. In practical SAGIN environments, CSI typically exhibits multimodal or irregular distributions owing to complex conditions. Forcing such patterns into a singular convex set inevitably introduces substantial geometric redundancy, which in turn leads to unnecessary resource consumption. Therefore, designing an adaptive uncertainty set that can effectively eliminate these redundant regions remains a critical challenge.

\section{\uppercase{{\large S}ystem {\large M}odel}}
\begin{figure}[!t]
    \centering
    \includegraphics[width=1.00\columnwidth]{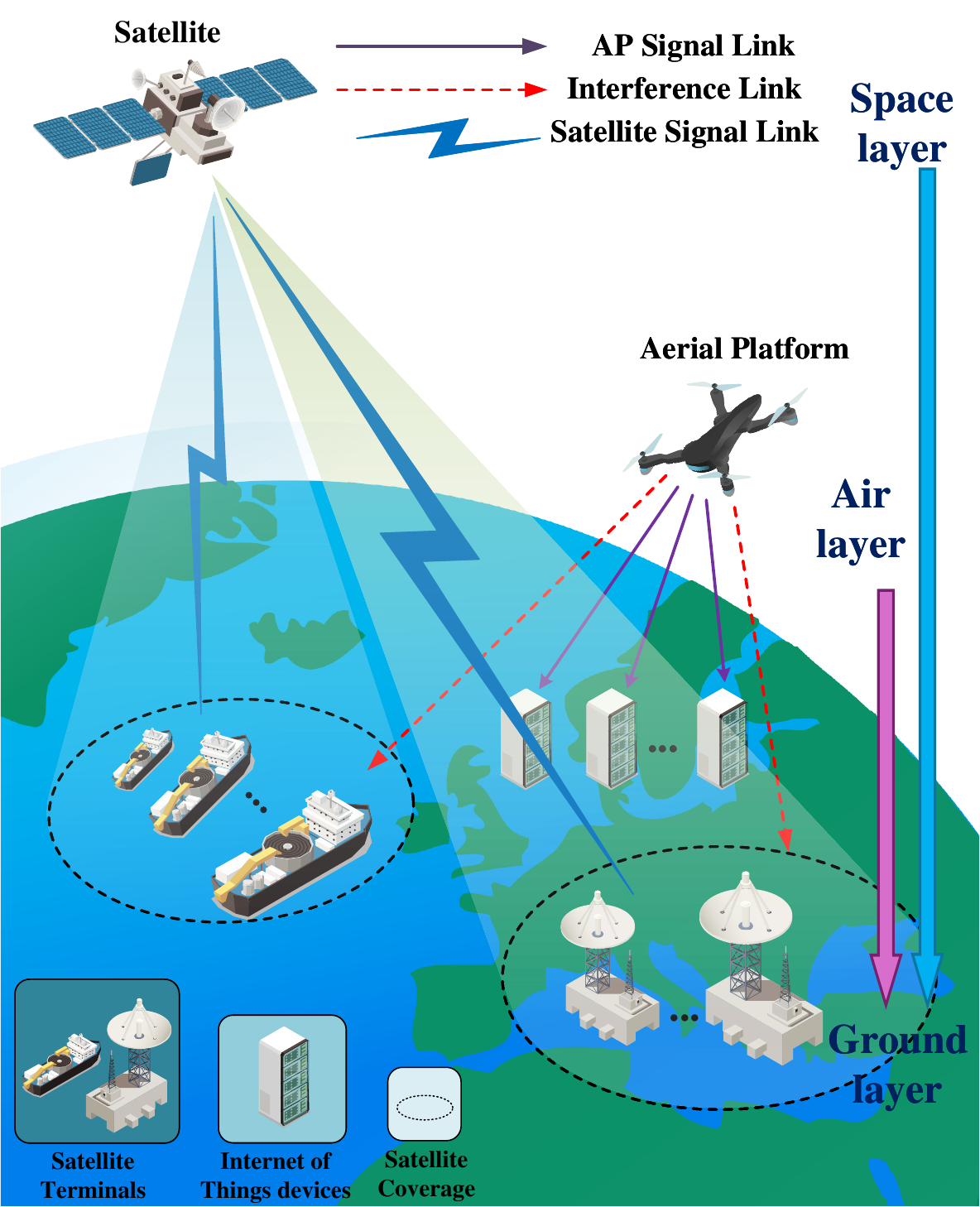}
    \caption{System model}
    \label{fig:scene}
\end{figure}
In this paper, we focus on the architecture of a SAGIN as depicted in Fig. \ref{fig:scene}. This heterogeneous network integrates a low earth orbit (LEO) satellite, the AP, STs, and IoTDs. The LEO satellite is equipped with an array-fed reflectarray antenna consisting of $N_s$ feed elements, which generates multiple directional beams to serve various STs simultaneously. Because STs possess robust communication capabilities and substantial antenna gains, they establish direct data links with the satellite. To further enhance the spectral efficiency of the satellite links, NOMA technology is employed, allowing the satellite to serve multiple STs within the same beam concurrently by superimposing their signals in the power domain.

Conversely, IoTDs are typically constrained by compact physical size, small antennas, and limited battery life. The severe path loss inherent in long-distance space-to-ground transmissions makes direct satellite communication highly inefficient and impractical for these low-power terminals. To bridge this connectivity gap, the AP is deployed at low altitudes to offer reliable, energy-efficient coverage to dense clusters of ground IoTDs. We assume that the AP is equipped with a uniform planar array (UPA) comprising $N_u$ antenna elements to concurrently serve $K$ active IoTDs. To efficiently meet diverse service demands, the AP adopts LDM technology. This enables the AP to broadcast common multicast streams (e.g., system-wide firmware updates or control messages) while simultaneously delivering specific unicast data to individual IoTDs over the shared frequency band. To maximize the overall spectral utilization of the SAGIN, we assume a full frequency reuse scheme where the satellite and the AP operate over the same downlink spectrum. Consequently, this spectrum sharing inevitably introduces cross-tier interference. Given the high transmit power of the AP and the strong reception capabilities of the STs, the interference generated by the AP toward the STs is significant and must be explicitly suppressed through precise beamforming. In contrast, the interference from the satellite to the IoTDs is assumed to be negligible, owing to the immense spatial propagation distance and the minimal reception gain of the IoTDs.
\subsection{Channel Model}
For accurate modeling of the satellite channel, we take into account the impacts of rain attenuation, free-space path loss, and the gain of both the on-board satellite antenna and the terrestrial user terminal antenna \cite{9610022}. On this basis, we formulate the channel vector (CV) from the satellite to the $n$-th ST user as:
\begin{equation}
    \mathbf{g}=\sqrt{G}\bm{\zeta}^{-\frac{1}{2}}\odot\mathbf{b}^{\frac{1}{2}}\left(\theta,\varphi\right)\odot{\hat{\mathbf{g}}}, \label{eq:satellite channel gain}
\end{equation}
where $\theta \in [0,\pi/2)$ and $\varphi \in [0,2\pi)$ denote the elevation and azimuth angles, respectively. $\odot$ denotes the Hadamard product, $G$ denotes the antenna gain of ST; $\bm{\zeta}\in\mathbb{C}^{N_s\times1}$ denotes the rain attenuation vector, where $\mathbb{C}^{N_s\times1}$ denotes the set of all $N_s\times1$ complex-valued column vectors. Furthermore, $\mathbf{b}\in\mathbb{C}^{N_s\times1}$ is introduced as the vector representing the gain of the satellite antenna, with its $i$-th element specified in
\begin{equation}
    b_{i}=b_{max}\left(\frac{J_1\left(u_{i}\right)}{2u_{i}}+36\frac{J_3\left(u_{i}\right)}{u_{i}^3}\right)^2 ,
\end{equation}
where $b_{max}$ denotes the maximum satellite beam gain. $J_1(\cdot)$ and $J_3(\cdot)$ represent the first-kind Bessel function of order 1 and 3, respectively. The parameter $u_{i}$ is defined as $u_{i}=2.07123\sin{\widetilde{\theta}}_{i}/\sin\theta_{3dB}$, where $\widetilde{\theta}_{i}$ denotes the angle between ST's position $\left(\theta,\varphi\right)$ and the $i$-th beam center $\left(\theta_i,\varphi_i\right)$ with respect to the satellite, and $\theta_{3dB}$ is the half-power beamwidth. Finally ${\hat{\mathbf{g}}}$ in Eq. (\ref{eq:satellite channel gain}) represents the channel response vector, with its $i$-th element modeled as
\begin{equation}
    {\hat{g}}_{i}=\frac{c}{4\pi f_cd_{i}}e^{-j\frac{2\pi f_cd_{i}}{c}},
\end{equation}
where $c$ represents the light speed, $f_c$ denotes the frequency of the carrier signal and $d_{i}$ represents the separation distance from ST to the $i$-th satellite antenna. 

Furthermore, given that the AP uses UPA to achieve substantial gain within a confined space, the CV $\mathbf{h}_k\in\mathbb{C}^{N_u\times1}$ for the $k$-th user served by the AP can be formulated as follows:
\begin{align}
    \mathbf{h}_k=&\sqrt{g_c\left(\phi_k,\psi_k\right)}\rho_0\mathbf{a}_x\left(\phi_k,\psi_k\right)\otimes\mathbf{a}_y\left(\phi_k,\psi_k\right) \nonumber\\
    &+\sqrt{\frac{1}{L_n}}\sum_{i=1}^{L_n}\sqrt{g_c\left(\phi_{k,i},\mathbf{\psi}_{k,i}\right)}\rho_i\mathbf{a}_x\left(\phi_{k,i},\mathbf{\psi}_{k,i}\right) \nonumber\\
    &\otimes\mathbf{a}_y\left(\phi_{k,i},\mathbf{\psi}_{k,i}\right),
    \label{eq:AP_CSI}
\end{align}
where $\phi_k\in[0,\pi/2)$ is the elevation angle, $\psi_k\in[0,2\pi)$ is the azimuth angle, $\otimes$ denotes the Kronecker product, and $L_n$ is the number of Non-Line-of-Sight (NLoS) paths. The parameters $\rho_0$ and $\rho_i$ represent the propagation losses for the Line-of-Sight (LoS) component and the $i$-th NLoS component, respectively. Based on the framework proposed by the ITU, the directional characteristic $g_c\left(\phi_k,\psi_k\right)$, expressed in dB and designated as ${\widetilde{g}}_c\left(\phi_k,\psi_k\right) = 10\log_{10}\left(g_c\left(\phi_k,\psi_k\right)\right)$, is formulated as follows:
\begin{equation}
{\widetilde{g}}_c\left(\phi_k,\psi_k\right)=G_{max}-\min\left\{g_x\left(\phi_k,\psi_k\right)+g_y\left(\phi_k,\psi_k\right),S_g\right\},
\end{equation}
where $G_{max}$ is the maximal antenna gain, while $S_g$ signifies the gain of the side lobes. Furthermore, $g_x\left(\phi_k,\psi_k\right)$ and $g_y\left(\phi_k,\psi_k\right)$ correspond to the normalized patterns along the X-axis and Y-axis, respectively, and they can be formulated as follows:
\begin{equation}
    g_x\left(\phi_k,\psi_k\right)=\min\left\{12\left(\frac{\arctan\left(\cot\phi_k/\cos\psi_k\right)}{\psi_x^{3dB}}\right)^2,S_g\right\},
\end{equation}
\begin{equation}
    g_y\left(\phi_k,\psi_k\right)=\min\left\{12\left(\frac{\arctan\left(\tan\phi_k\sin\psi_k\right)}{\psi_y^{3dB}}\right)^2,S_g\right\},
\end{equation}
where $\psi_x^{3dB}$ and $\psi_y^{3dB}$ are the 3dB beamwidths associated with the X-axis and Y-axis, respectively. In Eq. \eqref{eq:AP_CSI}, $\mathbf{a}_x\left(\phi_k,\psi_k\right)$ and $\mathbf{a}_y\left(\phi_k,\psi_k\right)$ signify the array steering vectors of the UPA for the X-axis and Y-axis directions, respectively, and they can be expressed as follows:
\begin{equation}
    \mathbf{a}_x=\left[1,e^{j\beta d_1\sin\phi_k\cos\psi_k},\cdots,e^{j\beta d_1\left(N_1-1\right)\sin\phi_k\cos\psi_k}\right]^T,
\end{equation}
\begin{equation}
    \mathbf{a}_y=\left[1,e^{j\beta d_2\sin\phi_k\sin\psi_k},\cdots,e^{j\beta d_2\left(N_2-1\right)\sin\phi_k\sin\psi_k}\right]^T,
\end{equation}
where $\beta=2\pi/\lambda_w$ represents the wave number. Finally, $N_1$ and $N_2$ denote the numbers of antennas that are uniformly distributed along the X-axis and Y-axis, respectively, with inter-element spacings of $d_1$ and $d_2$.

In our considered heterogeneous communication network, the AP hovers quasi-statically above ground IoT terminals, featuring short transmission ranges and dominant LoS paths. Meanwhile, the low mobility of massive IoTDs allows the AP to perform frequent pilot-based channel estimation and capture  high-precision CSI with small estimation errors. Unlike the AP, acquiring accurate CSI for satellite links is severely hindered by the large RTT and significant Doppler shifts caused by the high-speed motion of the satellite and terminals. Therefore, we assume that perfect CSI is available only for the AP, while the satellite links are subject to CSI uncertainty.
Specifically, based on the satellite channel physical model derived in Eq. \eqref{eq:satellite channel gain}, let $\mathbf{g}_{m,n}$ denote the true channel vector corresponding to the $n$-th ST served by the $m$-th beam. We characterize the imperfection of the acquired CSI via the following additive error formulation:
\begin{equation}
\mathbf{g}_{m,n}=\tilde{\mathbf{g}}_{m,n}+\mathbf{e}_{m,n},
\end{equation}
where $\tilde{\mathbf{g}}_{m,n}$ is the estimated channel vector, and $\mathbf{e}_{m,n}$ captures the estimation error arising from channel aging and measurement inaccuracies.
\subsection{Signal Model}
In our proposed SAGIN downlink transmission, we assume that all transmitted information symbols are normalized in power. Let $s_{m,n}(t)$ be the intended symbol for the $n$-th ST in the $m$-th beam, and $\alpha_{m,n}\in[0, 1]$ represent its assigned power allocation coefficient, where $\sum_{n=1}^{N_{m}} \alpha_{m,n} = 1$ for all $m$. The aggregate signal transmitted by the satellite is formulated as
\begin{equation}
\mathbf{x}_{s}(t) = \sum_{m=1}^{M} \mathbf{v}_{m} \sum_{n=1}^{N_{m}} \sqrt{\alpha_{m,n}} s_{m,n}(t),
\end{equation}
where $\mathbf{v}_{m}\in\mathbb{C}^{N_s\times1}$ is beamforming weight vectors.

The AP employs the LDM scheme to broadcast a common multicast stream $s_0(t)$ to all $K$ associated IoTDs while delivering dedicated unicast streams $s_k(t)$ to the $k$-th IoTD. Let $\mathbf{w}_0 \in \mathbb{C}^{N_u \times 1}$ and $\mathbf{w}_k \in \mathbb{C}^{N_u \times 1}$ denote their respective beamforming vectors. Then, the total transmitted signal from the AP can be represented as
\begin{equation}
    \mathbf{x}_{u}(t) = \mathbf{w}_{0} s_{0}(t) + \sum_{k=1}^{K} \mathbf{w}_{k} s_{k}(t).
\end{equation}

To extract the desired information from the superposed signals, appropriate decoding strategies are performed at the receivers. At the ST side, receivers for users within the same beam apply successive interference cancellation (SIC) based on NOMA principles. Assuming the STs are sorted in ascending order of their effective channel qualities, the $n$-th ST first decodes and cancels the interference from users allocated with more power (i.e., users $j < n$), while treating the channel conditions of stronger users as background noise (i.e., users $j > n$). Consequently, the resulting SINR for the $(m,n)$-th ST is formulated as
\begin{equation}
\gamma_{m,n} = \frac{\alpha_{m,n}\left|\mathbf{g}_{m,n}^H\mathbf{v}_m\right|^2}{
    \begin{aligned}
        & \bigg(\underbrace{\sum_{j=n+1}^{N_{m}}\alpha_{m,j}\left|\mathbf{g}_{m,n}^H\mathbf{v}_m\right|^2}_{\text{intra-beam interference}} +
          \underbrace{\sum_{i\neq m}^{M}\left|\mathbf{g}_{m,n}^H\mathbf{v}_i\right|^2}_{\text{inter-beam interference}} \\
        & + \underbrace{\sum_{k=0}^{K}\left|\mathbf{h}_{m,n}^H\mathbf{w}_k\right|^2}_{\text{AP interference}} +
          \underbrace{\sigma_{m,n}^2}_{\text{noise}}\bigg)\\
        &\forall m, \forall n,
    \end{aligned}
}\label{eq:satellite SINR}
\end{equation}
where $\sigma^2$ denotes the power of additive white Gaussian noise (AWGN). In addition, $\mathbf{h}_{m,n}$ denote the CV of AP to the $n$-th ST in the $m$-th beam.
It is worth noting that due to the high transmit power of the AP and the robust receive gain of the STs, both the multicast ($k=0$) and unicast ($k \ge 1$) transmissions from the AP inflict considerable cross-tier interference on the STs. Conversely, for the $k$-th IoTD, the cross-tier interference originating from the satellite is considered practically negligible owing to the substantial free-space path loss and the limited antenna gain of the IoTD device.

At the IoTD side, a two-layer decoding mechanism is implemented in accordance with the LDM protocol. The IoTD first directly decodes the common multicast stream $s_0(t)$, treating all concurrent unicast transmissions as interference. Subsequently, it subtracts the reconstructed multicast signal from the total received signal $y_k(t)$ to decode its dedicated unicast stream $s_k(t)$. Based on this sequential decoding framework, the SINR for the multicast and unicast signals at the $k$-th IoTD are respectively calculated as
\begin{equation}
    \gamma_{k}^{mc} = \frac{|\mathbf{h}_{k}^H\mathbf{w}_0|^2}{\sum_{i=1}^{K}|\mathbf{h}_{k}^H\mathbf{w}_i|^2 + \sigma_{k}^2},
\end{equation}
\begin{equation}
    \gamma_{k}^{uc}=\frac{\left|\mathbf{h}_{k}^H\mathbf{w}_k\right|^2}{\sum_{\substack{i=1 \\ i\neq k}}^{K}\left|\mathbf{h}_{k}^H\mathbf{w}_i\right|^2+\sigma_{k}^2}.
\end{equation}

\subsection{Problem Formulation and simplification}
Considering the heterogeneous nature of the network, we assume that STs and IoTDs impose distinct QoS requirements. This study focuses on the optimization of beamforming vectors serving the STs and IoTDs, subject to individual minimum SINR constraints. Furthermore, acknowledging the inevitability of CSI uncertainty, we assume that these terminal groups can tolerate a specific maximum outage probability (i.e., probability of communication interruption). Based on these premises, the robust beamforming optimization problem is formulated as follows:
\begin{subequations}
    \label{eq:problem formulation}
    \begin{align}
        &\min_{\{\mathbf{v}_m\},\{\mathbf{w}_k\}, \{\alpha_{m,n}\}}\sum_{m = 1}^{M}\Vert\mathbf{v}_m\Vert^2 + \sum_{k = 0}^{K}\Vert\mathbf{w}_k\Vert^2 \label{eq:pf-a}\\
        \mathrm{s.t.}\quad&\Pr\left\{\gamma_{m,n} \geq \Gamma_{m,n}\right\} \geq 1 - \epsilon,\forall m, \forall n, \label{eq:PF-b} \\
        &\gamma_{k}^{mc} \geq \Gamma_{k}^{mc},\forall k,   \label{eq:pf-c}\\
        &\gamma_{k}^{uc} \geq\Gamma_{k}^{uc}, \forall k,   \label{eq:pf-d}\\
        &\sum_{n=1}^{N}\alpha_{m,n}=1,0 < \alpha_{m,n} < 1,\forall m,\forall n,\label{eq:pf-e}
    \end{align}
\end{subequations}
where $\Gamma_{m,n}$ represents the QoS threshold of $\left(m,n\right)$-th ST, $\Gamma_k^{mc}$ and $\Gamma_k^{uc}$ represent the multicast and unicast QoS threshold of $k$-th IoTD, respectively. $\epsilon$ denotes the maximum outage probability that users can tolerate during communication.

Directly solving this problem faces two core challenges: First, the chance constraints in (\ref{eq:problem formulation}b) make the problem a semi-infinite optimization problem that is intractable for standard convex solvers; Second, the coupled optimization variables and non-convex constraints further increase the difficulty of obtaining the optimal solution. To address these challenges, we first perform equivalent transformation and convex relaxation on the original problem, and prove that the robust constraints based on the uncertainty set serve as a sufficient condition for the chance constraints, laying a theoretical foundation for the subsequent DNN-driven robust optimization framework.

To facilitate the subsequent derivation and solution, we first perform equivalent transformation on the SINR constraints and the objective function in the original problem.
First, for the convenience of analysis, let $\mathbf{V}_m=\mathbf{v}_m\mathbf{v}_m^H$, $\mathbf{W}_k=\mathbf{w}_k\mathbf{w}_k^H$, $\mathbf{C}_{m,n} = \mathbf{g}_{m,n}\mathbf{g}_{m,n}^H$, $\mathbf{H}_{m,n} = \mathbf{h}_{m,n}\mathbf{h}_{m,n}^H$, $\mathbf{H}_{k} = \mathbf{h}_{k}\mathbf{h}_{k}^H$ and $\tilde{\alpha}_{m,n} = \alpha_{m,n} - \Gamma_{m,n}\sum_{j=n+1}^{N}\alpha_{m,j}$ then the inequalities $\gamma_{m,n} \geq \Gamma_{m,n}$, $\gamma_k^{mc} \geq \Gamma_k^{mc}$ and $\gamma_k^{uc} \geq \Gamma_k^{uc}$ can be reformulated as
\begin{equation}
    \begin{aligned}
        &\mathrm{Tr}(\tilde{\alpha}_{m,n}\mathbf{C}_{m,n}\mathbf{V}_m) - \Gamma_{m,n}\sum_{\substack{i=1 \\ i\neq m}}^{M}\mathrm{Tr}(\mathbf{C}_{m,n}\mathbf{V}_i)\\
        & - \Gamma_{m,n}\sum_{k=0}^{K}\mathrm{Tr}(\mathbf{H}_{m,n}\mathbf{W}_k) -\Gamma_{m,n}\sigma_{m,n}^2 \geq 0,
    \end{aligned}
    \label{eq:simplify ST}
\end{equation}
\begin{equation}
    \mathrm{Tr}(\mathbf{H}_{k}\mathbf{W}_0) - \Gamma_{k}^{mc}(\sum_{i=1}^K \mathrm{Tr}(\mathbf{H}_{k}\mathbf{W}_i) + \sigma_k^2) \geq 0,
    \label{eq:simplify AP mc}
\end{equation}
\begin{equation}
    \mathrm{Tr}(\mathbf{H}_{k}\mathbf{W}_k) - \Gamma_{k}^{uc}(\sum_{\substack{i=1 \\ i\neq k}}^K \mathrm{Tr}(\mathbf{H}_{k}\mathbf{W}_i) + \sigma_k^2) \geq 0.
   \label{eq:simplify AP nc}
\end{equation}
The left-hand sides of inequalities (\ref{eq:simplify ST}), (\ref{eq:simplify AP mc}) and (\ref{eq:simplify AP nc}) can be represented by $F_{ST}\left(\alpha_{m,n}, \mathbf{g}_{m,n}, \mathbf{V}, \mathbf{W}\right)$, $F_{mc}\left(\mathbf{h}_k,\mathbf{W}\right)$ and $F_{uc}\left(\mathbf{h}_k,\mathbf{W}\right)$, i.e.,
\begin{equation}
    F_{ST}\left(\alpha_{m,n}, \mathbf{g}_{m,n}, \mathbf{V}, \mathbf{W}\right) \geq 0,
\end{equation}
\begin{equation}
    F_{mc}\left(\mathbf{h}_k,\mathbf{W}\right) \geq 0,
\end{equation}
\begin{equation}
    F_{uc}\left(\mathbf{h}_k,\mathbf{W}\right) \geq 0.
\end{equation}
After the above processes, Problem (\ref{eq:problem formulation}) can be reformulated as
\begin{subequations}
    \label{eq:transformed-1 problem formulation} 
    \begin{align}   
        &\min_{\left\{\mathbf{V}_m\right\}\succeq0,\left\{\mathbf{W}_k\right\}\succeq0,\{\alpha_{m,n}\} }\sum_{m = 1}^{M}\mathrm{Tr}(\mathbf{V}_m)+\sum_{k = 0}^{K}\mathrm{Tr}(\mathbf{W}_k) \label{eq:t1pf-a} \\
        \mathrm{s.t.}\quad&\Pr\left\{F_{ST}\left(\alpha_{m,n}, \mathbf{g}_{m,n}, \mathbf{V}, \mathbf{W}\right) \geq 0\right\}\geq 1 - \epsilon, \forall m, \forall n,\label{eq:t1pf-b} \\
        &F_{mc}\left(\mathbf{h}_k,\mathbf{W}\right) \geq 0, \forall k , \label{eq:t1pf-c}\\
        &F_{uc}\left(\mathbf{h}_k,\mathbf{W}\right) \geq 0,\forall k , \label{eq:t1pf-d}\\
        &\mathrm{rank}\left(\mathbf{V}_m\right)=1, \forall m, \label{eq:t1pf-e}\\
        &\mathrm{rank}\left(\mathbf{W}_k\right)=1, \forall k, \label{eq:t1pf-f}\\
        &\sum_{n=1}^{N}\alpha_{m,n}=1,0 < \alpha_{m,n} < 1,\forall m,\forall n. \label{eq:t1pf-g}
    \end{align}
\end{subequations}

Solving Problem (\ref{eq:transformed-1 problem formulation}) directly is still difficult due to the chance constraints. To address this, we reformulate the problem using uncertainty sets, denoted as $\mathcal{U}_{m,n}$. We design these sets to capture the channel vectors $\mathbf{g}_{m,n}$ with a confidence level of $1-\epsilon$. By ensuring that the constraints $F_{ST}\left(\alpha_{m,n}, \mathbf{g}_{m,n}, \mathbf{V}, \mathbf{W}\right) \geq 0$ hold for any point within these sets, we can guarantee the satisfaction of the chance constraints. Hence, the critical task is to accurately define the geometry of $\mathcal{U}_{m,n}$.

Based on the above motivations, we reformulate Eq. (\ref{eq:t1pf-b}) as follows:
\begin{equation}
    \label{eq:transformed-2 problem formulation} 
        F_{ST}\left(\alpha_{m,n}, \mathbf{g}_{m,n}, \mathbf{V}, \mathbf{W}\right) \geq 0, \forall \mathbf{g}_{m,n}\in \mathcal{U}_{m,n},\forall m, \forall n,
\end{equation}

Although SDR is known to be tight for conventional MISO-NOMA beamforming~\cite{8959161}, this result cannot be invoked verbatim in our setting. We therefore provide a self-contained analysis in Section~\ref{section:Formation}. On this basis, we drop the rank-one constraints and reformulate the Problem \eqref{eq:transformed-1 problem formulation} as follows:
\begin{subequations}
    \label{eq:transformed-3 problem formulation} 
    \begin{align}   
        &\min_{\left\{\mathbf{V}_m\right\}\succeq0,\left\{\mathbf{W}_k\right\}\succeq0,\{\alpha_{m,n}\} }\sum_{m = 1}^{M}\mathrm{Tr}(\mathbf{V}_m)+\sum_{k = 0}^{K}\mathrm{Tr}(\mathbf{W}_k) \label{eq:t3pf-a} \\
        \mathrm{s.t.}\quad&\eqref{eq:transformed-2 problem formulation},\eqref{eq:t1pf-c}, \eqref{eq:t1pf-d},\eqref{eq:t1pf-g}. \label{eq:t3pf-b}
    \end{align}
\end{subequations}

After the intractable chance constraints and non-convex objective transformed into the deterministic robust convex optimization form in Problem (\ref{eq:transformed-3 problem formulation}), the remaining core challenge is to precisely construct a compact, distribution-matched uncertainty set $\mathcal{U}_{m,n}$. To address this, we propose a DNN-driven robust resource management framework to realize uncertainty set modeling and joint beamforming-resource allocation optimization, which is detailed in the following section.
\section{\uppercase{{\large D}nn-{\large D}riven {\large R}obust {\large R}esource  {\large M}anagement {\large F}ramework } }
\begin{figure*}[htbp]
    \centering
    \includegraphics[width=1.9\columnwidth]{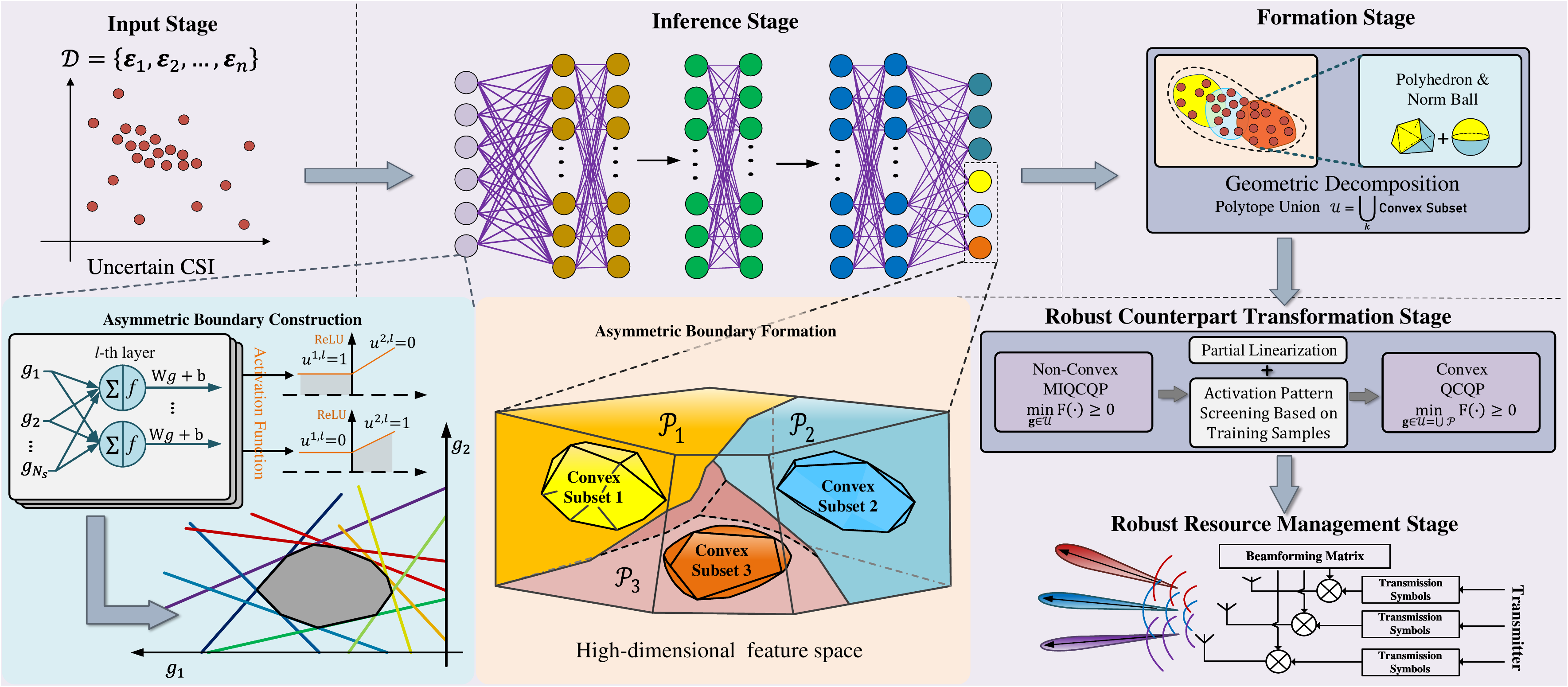}
    \caption{Architecture of the proposed DNN-driven robust resource management framework.}
    \label{fig:DNN_Algorithm_framework}
\end{figure*}
In this section, we propose a DNN-driven robust resource management framework to solve the formulated chance-constrained optimization problem. The overall architecture of the proposed framework is illustrated in detail in Fig. \ref{fig:DNN_Algorithm_framework}. As depicted in the figure, the framework consists fundamentally of four interconnected stages: Input and Inference Stage, Uncertainty Set Formation Stage, Robust Counterpart Transformation Stage, and Robust Resource Management Stage. First, historical CSI samples are collected and fed into a pre-trained DNN. By leveraging the feature learning capabilities of DNN, the Inference Stage extracts the high-dimensional features of the channel uncertainties. Subsequently, in the Formation Stage, the global nonlinear mapping of the DNN is mathematically decomposed to construct an asymmetric, tightly fitted uncertainty set. On this basis, the Robust Counterpart Transformation Stage converts the intractable constraints into a sequence of solvable convex constraints. Finally, in the Robust Resource Management Stage, we design a two-layer iterative algorithm to jointly optimize the beamforming matrices and power allocation coefficients. The detailed mathematical transformations and algorithmic implementations corresponding to these four stages are elaborated in Section IV-A, IV-B, IV-C, and IV-D, respectively.

\subsection{Input and Inference Stage for CSI Uncertainty}
This stage learns and infers uncertain CSI features via the DNN to construct the uncertainty set. First, to model the statistical distribution of CSI uncertainty, we collect $D$ independent and identically distributed (i.i.d.) historical samples of the actual CV $\mathbf{g}_{m,n}$, to construct the training dataset, which is denoted as: $$\mathcal{D}_{m,n} = \{\boldsymbol{\varepsilon}_1^{m,n}, \boldsymbol{\varepsilon}_2^{m,n}, \cdots, \boldsymbol{\varepsilon}_D^{m,n}\}.$$
With the dataset $\mathcal{D}$ properly acquired, instead of traditional geometric sets, we employ a data-driven approach by training a DNN to learn the hidden structures within the CSI data. The entire computational process of the neural network can be expressed as
\begin{equation}
    f_{\mathbf{T}^1,\dots,\mathbf{T}^L}\left(\bm{\varepsilon}\right)=\delta^L\left(\mathbf{T}^L\delta^{L-1}\left(\mathbf{T}^{L-1}\dots\delta^1\left(\mathbf{T}^1\bm{\varepsilon}\right)\right)\right),
\end{equation}
where $\mathbf{T}^l \in \mathbb{R}^{d_l \times d_{l-1}}$ is the weight matrice, and $\delta^l : \mathbb{R} \rightarrow \mathbb{R}$ for each $l \in [L]$ is activation function which is applied component-wise. Here, $[L]=\{1,2,\ldots,L\}$ denotes the set of all layer indices in the DNN. The dimension $d_l$ is called the width of the $l$-th layer with $d_0 := \text{dim}\left(\bm{\varepsilon}\right)$.

We assume that all activation functions are continuous piecewise affine functions, i.e., they are of the form
\begin{equation}
    \delta^l\left(w\right)=a_i^lw+b_i^l \quad \text{if} \quad \underline{\tau}^{i,l} \leq w < \overline{\tau}^{i,l}, \forall i \in [Z_l],
    \label{eq:activation function}
\end{equation}
for all $l\in[L]$, where $a_i^l \in\mathbb{R}$ are the given slopes, $b_i^l$ are the given intercept on the y-axis, $Z_l \in \mathbb{N} $ is the number of intervals, and $[Z_l] = \{1,2,\dots,Z_l\}$ denotes the set of all interval indices for the $l$-th layer. Moreover, $\underline{\tau}^{i,l} \leq \overline{\tau}^{i,l}$ are the interval bounds that satisfy $\overline{\tau}^{i,l}=\underline{\tau}^{i+1,l}$, $\underline{\tau}^{1,l}=-\infty$ and $\underline{\tau}^{Z_l,l}=\infty$  for all $l\in[L]$. Geometrically, the boundaries of these continuous piecewise affine functions act as a set of intersecting hyperplanes that partition the high-dimensional feature space into distinct polyhedral regions, as illustrated in the ``Asymmetric Boundary Construction" of Fig. \ref{fig:DNN_Algorithm_framework}. Note that the rectified linear unit (ReLU) activation function $\delta\left(w\right) = \max\left\{w,0\right\}$ can be modeled from Eq. (\ref{eq:activation function}) by $Z_l=2$, $a_1^l=0$, $a_2^l=1$, $b_1^l=b_2^l=0$, $\underline{\tau}^{1,l}=-\infty$, $\overline{\tau}^{1,l}=0$, $\underline{\tau}^{2,l}=0$ and $\overline{\tau}^{2,l}=\infty$. Other piecewise affine activation functions such as the Hardtanh or the hard sigmoid function can be modeled from Eq. (\ref{eq:activation function}) as well. In general, any continuous function could be approximated by these piecewise affine functions.

To train the DNN for accurate CSI uncertainty characterization, we adopt a quantile-based optimization objective tailored to the outage probability constraint in our problem. The core training goal is to learn a nonlinear feature mapping that tightly encloses the majority of normal CSI samples while excluding outliers, with the coverage rate strictly aligned to the preset confidence level $1-\epsilon$.

For the training dataset $\mathcal{D}_{m,n}=\left\{ \boldsymbol{\varepsilon}_{1}^{m,n},\boldsymbol{\varepsilon}_{2}^{m,n},\cdots ,\boldsymbol{\varepsilon}_{D}^{m,n}\right\}$, we first perform a forward pass on all samples through the DNN to obtain their feature representations in the output layer, and calculate the Euclidean distance between each sample's feature representation and the pre-defined center $\bar{c}$, denoted as the radius of each sample:
\begin{equation}
r_i = \left\| f_{\mathbf{T}^{1},\dots,\mathbf{T}^{L}}(\boldsymbol{\varepsilon}_i^{m,n})-\overline{c} \right\|_2, \quad \forall i \in [D].
\end{equation}

We then sort all sample radii in ascending order to obtain the sorted sequence $r_{\pi(1)} \leq r_{\pi(2)} \leq \dots \leq r_{\pi(D)}$, where $\pi(\cdot)$ represents the index mapping after sorting. To enforce the network to satisfy the $1-\epsilon$ coverage probability requirement, we design a weighted quantile loss function that focuses on the samples around the $(1-\epsilon)$-quantile of the radius distribution:
\begin{equation}
\mathcal{L} = \sum_{i=1}^{t} \mu _i \cdot r_{\pi(\lfloor (1-\epsilon)D \rfloor -i)} - \sum_{i=1}^{t} \nu _i \cdot r_{\pi(\lfloor (1-\epsilon)D \rfloor +i)}.
\end{equation}

In this loss function, $\lfloor \cdot \rfloor$ denotes the floor operation, and $t$ is the number of adjacent samples around the target quantile that participate in the loss calculation. The positive weight coefficients $\mu _i$ and $\nu _i$ control the gradient magnitude of the loss function: a larger $\mu _i$ penalizes the shrinkage of the uncertainty set boundary (to avoid under-coverage of valid CSI samples), while a larger $\nu _i$ penalizes the expansion of the boundary (to eliminate redundant geometric space and reduce conservatism). 


After the quantile-based training, we obtain the fixed optimal weight matrices $\mathbf{T}=\left(\mathbf{T}^1,\dots,\mathbf{T}^L\right)$, a center point $\bar{c} \in \mathbb{R}^{d_L}$ and a radius $R>0$, which establishes a nonlinear mapping from the raw CSI space to the high-dimensional feature space. Based on this well-trained mapping, we give the initial implicit formulation of the uncertainty set as:
\begin{equation}
    \mathcal{U}:=\left\{\mathbf{g} \left| \left\|f_{\mathbf{T}^1,\dots,\mathbf{T}^L}\left(\mathbf{g}\right)-\bar{c}\right\| \leq R \right. \right\}.
    \label{eq:uncertainty}
\end{equation}
It is important to note that the center point $\bar{c}$ is fixed and is not considered a decision variable. A commonly adopted strategy is to set $\bar{c}$ as the average of the network representations obtained from performing an initial forward pass on the training data samples. The radius $R$ is determined by the preset outage probability threshold $\epsilon$, i.e., $R=r_{\pi(1-\epsilon)D}$, which ensures that the probability of the CSI samples falling into the set $\mathcal{U}$ satisfies $\Pr\{\mathbf{g} \in \mathcal{U}\} \geq 1-\epsilon$, which is consistent with the chance constraint in the original problem. Once the network representations are inferred from the training samples, we need to decompose the global nonlinear mapping of the DNN, i.e., Eq. \eqref{eq:uncertainty},  to form explicit boundaries of the uncertainty set, which is the core content of the next section.
\subsection{Formation Stage of Asymmetric Boundaries}
Although the neural network $f_{{\mathbf{T}^1,\dots,\mathbf{T}^L}}\left(\cdot\right)$ is globally non-linear and non-convex, the use of piecewise affine activation functions allows us to characterize the uncertainty set $\mathcal{U}(\mathbf{T},\bar{c},R)$ as a union of tractable convex sets. To demonstrate this, we first introduce the concept of an activation pattern.

For any given CV $\mathbf{g}$, the output of each neuron in layer $l$ falls into exactly one of the $Z_l$ intervals defined in Eq. (\ref{eq:activation function}). We can encode these decisions using binary variables. Let $\mathcal{B}$ denote the set of all feasible activation patterns, defined as
\begin{equation}
\!\!\!\mathcal{B}\!:=\!\left\{\!\mathbf{u}\!:=\!(\mathbf{u}^{i,l})_{\substack{l\in[L]\\i\in[Z_l]}} \bigg| \mathbf{u}^{i,l}\!\!\in\{0,1\}^{d_l}\!, \!\sum_{i=1}^{Z_l}\mathbf{u}^{i,l}\!=\!\mathbf{1}_{d_l \times 1}\!\right\},
\end{equation}
where the vectors $\mathbf{u}$ encode the activation decisions of all neurons, i.e., $u_j^{i,l} = 1$ if the outcome of the j-th component of layer $l$ lies in the interval $\left[\underline{\tau}^{i,l}, \overline{\tau}^{i,l}\right)$, and 0 otherwise, the constraint $\sum_{i=1}^{Z_l} \mathbf{u}^{i,l} = \mathbf{1}_{d_l \times 1}$ ensures that exactly one interval is chosen for each neuron. We refer to a vector $\mathbf{u} \in \mathcal{B}$ as an activation pattern.

Consider a fixed activation pattern $\mathbf{u} \in \mathcal{B}$, the activation function $\delta^l(w)$ degenerates into a simple linear transformation $a_i^l w + b_i^l$. Consequently, the entire neural network collapses into a single affine transformation. By composing the affine functions of each layer, the network output for a fixed $\mathbf{u}$ can be expressed as
\begin{equation}
f_{\mathbf{T}}(\mathbf{g})|_\mathbf{u} = \tilde{\mathbf{T}}^{(L+1)}\mathbf{g} + \tilde{\mathbf{b}}^{(L+1)}, 
\end{equation}
where $\tilde{\mathbf{T}}^{(L+1)}$ and $\tilde{\mathbf{b}}^{(L+1)}$ are the composite weight matrix and the composite bias vector, respectively, which are derived recursively. Specifically, the layer-wise expression for $\tilde{\mathbf{T}}^{l}$ is given by:
\begin{equation}
\tilde{\mathbf{T}}^{l} = \left( \prod_{j=1}^{l-1} \mathbf{T}^{j+1} \text{diag}\left(\sum_{i=1}^{Z_j} \mathbf{u}^{i,j} a_i^j\right) \right) \mathbf{T}^1,\label{eq:weight matrix}
\end{equation}
with initial conditions $\tilde{\mathbf{T}}^1=\mathbf{T}^1$, $\mathbf{T}^{L+1}=\mathbf{I}_{d_L}$, where $\mathbf{I}_{d_L}$ is the identity matrix in dimension $d_L$. Similarly, initialized with $\tilde{\mathbf{b}}^1=\mathbf{0}$, the composite bias vector $\tilde{\mathbf{b}}^{l}$ is accumulated through the layers as:
\begin{equation}
    \tilde{\mathbf{b}}^l\!=\!\sum_{j=2}^{l}\mathbf{T}^l\!\!\left(\prod_{s=j}^{l-    1}\text{diag}\left(\sum_{i=1}^{Z_s}\mathbf{u}^{i,s}a_i^s\right)\mathbf{T}^s\right)\!\!\!\left(\sum_{i=1}^{Z_{j-1}}\mathbf{u}^{i,j-1}b_i^{j-1}\right)\!\!.\label{eq:bias vector}
\end{equation}

To activate a specific pattern $\mathbf{u}$, the CV $\mathbf{g}$ must satisfy the linear inequalities corresponding to the selected intervals at every layer. Combining these validity constraints with the uncertainty radius constraint, we define a convex set $\mathcal{P}(\mathbf{u})$ for each pattern $\mathbf{u}$:
\begin{equation}
    \mathcal{P}\left(\mathbf{u}\right) := \left\{ \mathbf{g} \left|
        \begin{aligned}
            & \tilde{\mathbf{T}}^{l}\mathbf{g} + \tilde{\mathbf{b}}^{l} < \sum_{i=1}^{Z_l} \mathbf{u}^{i,l} \overline{\tau}^{i,l},  \forall l \in [L], \\
            & \tilde{\mathbf{T}}^{l}\mathbf{g} + \tilde{\mathbf{b}}^{l} \geq \sum_{i=1}^{Z_l} \mathbf{u}^{i,l} \underline{\tau}^{i,l},  \forall l \in [L], \\
            & \left\|\tilde{\mathbf{T}}^{L+1}\mathbf{g}+ \tilde{\mathbf{b}}^{L+1} - \bar{c}\right\| \leq R
        \end{aligned}\right.\right\}.
\end{equation}
The first two sets of inequalities ensure that $\mathbf{g}$ is consistent with the activation pattern $\mathbf{u}$, while the final inequality imposes the radius constraint. Since every possible input vector $\mathbf{g}$ corresponds to a specific activation pattern, the total uncertainty set $\mathcal{U}$ is the union of these convex regions over all possible patterns in $\mathcal{B}$. Thus, we arrive at the following structural characterization:
\begin{equation}
    \mathcal{U} = \bigcup_{\mathbf{u} \in \mathcal{B}} \mathcal{P}(\mathbf{u}).
\end{equation}
This structural characterization reveals that the DNN-bounded uncertainty set is equivalent to a finite union of convex sets. As visually depicted in the ``Asymmetric Boundary Formation" and ``Formation Stage" blocks of Fig. \ref{fig:DNN_Algorithm_framework}, each subset $\mathcal{P}(\mathbf{u})$ represents a polyhedron intersected with a norm ball.

\subsection{Robust Counterpart Transformation Stage \label{section:Formation}}
With the uncertainty set explicitly formulated as a tractable union, the subsequent task is to embed these geometric boundaries into the constraints. However, Eq. \eqref{eq:transformed-2 problem formulation} contains quadratic terms in the CV $\mathbf{g}_{m,n}$. To handle these quadratic terms, we introduce the auxiliary matrix $\mathbf{G}_{m,n}=\mathbf{g}_{m,n}\mathbf{g}_{m,n}^H$, which satisfying the rank-one constraint $\mathrm{rank}(\mathbf{G}_{m,n})=1$ to transform the original non-convex function $F_{ST}$ into a linear function with respect to $\mathbf{G}_{m,n}$:
\begin{multline}
\tilde{F}_{ST}(\alpha_{m,n}, \mathbf{G}_{m,n}, \mathbf{V}, \mathbf{W})
\!=\! \mathrm{Tr}\big(\tilde{\alpha}_{m,n} \mathbf{G}_{m,n} \!\mathbf{V}_m\big) \!-\! \Gamma_{m,n}\sigma_{m,n}^2 \\
- \Gamma_{m,n}\sum_{\substack{i=1 \\ i\neq m}}^{M}\mathrm{Tr}\big(\mathbf{G}_{m,n}\mathbf{V}_i\big)
- \Gamma_{m,n}\sum_{k=0}^{K}\mathrm{Tr}\big(\mathbf{H}_{m,n}\mathbf{W}_k\big),
\end{multline}

Consequently, the constraint requires $\tilde{F}_{ST} \ge 0$ for any channel in the uncertainty set $\mathcal{U}_{m,n}$. To satisfy this constraint, we need to ensure that the minimum value of $\tilde{F}_{ST}$ within $\mathcal{U}_{m,n}$ remains non-negative. Identifying this channel realization constitutes the adversarial sub-problem (ASP), which is formulated as:
\begin{subequations}
    \begin{align}
        &\min_{\mathbf{G}_{m,n},\mathbf{g}_{m,n}}\tilde{F}_{ST}\left(\alpha_{m,n}, \mathbf{G}_{m,n}, \mathbf{V}, \mathbf{W}\right) \geq 0\\
        \mathrm{s.t.}\quad&\mathbf{g}_{m,n} \in \mathcal{U}_{m,n},\forall m,\forall n, \\
        &\mathbf{G}_{m,n}\succeq \mathbf{g}_{m,n}\mathbf{g}_{m,n}^H,\\
        &\mathrm{rank}(\mathbf{G}_{m,n})=1.
    \end{align}
    \label{eq:transformed-5 problem formulation}
\end{subequations}
As derived in the previous section, the set $\mathcal{U}$ is the union of convex regions $\mathcal{P}(\mathbf{u})$ corresponding to valid activation patterns $\mathbf{u} \in \mathcal{B}$. Therefore, optimizing over $\mathcal{U}$ involves simultaneously searching for the least favorable activation pattern $\mathbf{u}$ and the least favorable parameter $\mathbf{g}_{m,n}$ within that pattern. By introducing binary variables to represent the activation decisions $\mathbf{u}^{i,l} \in \{0,1\}^{d_l}$, the DNN collapses into a single affine transformation, denoted as $\mathbf{g}^{L+1} = \tilde{\mathbf{T}}_{\mathbf{u}} \mathbf{g} + \tilde{\mathbf{b}}_{\mathbf{u}}$, where $\tilde{\mathbf{T}}_{\mathbf{u}}$ and $\tilde{\mathbf{b}}_{\mathbf{u}}$ are the composite weight matrix and bias vector derived from Eq. \eqref{eq:weight matrix} and \eqref{eq:bias vector}. The radius constraint $\left\|\mathbf{g}^{L+1} - \overline{c}\right\|^2 \leq R^2$ can be equivalently expanded using the newly introduced matrix $\mathbf{G}_{m,n}$ as:
\begin{equation}
\text{Tr}(\tilde{\mathbf{T}}_{\mathbf{u}}^H \tilde{\mathbf{T}}_{\mathbf{u}} \mathbf{G}_{m,n}) + 2\text{Re}\{(\tilde{\mathbf{b}}_{\mathbf{u}} - \overline{c})^H \tilde{\mathbf{T}}_{\mathbf{u}} \mathbf{g}_{m,n}\} + \|\tilde{\mathbf{b}}_{\mathbf{u}} - \overline{c}\|^2 \leq R^2.\nonumber
\end{equation}
For this ASP, we adopt the SDR technique. By dropping the rank-one constraint and replacing it with a LMI via Schur complement, we can reformulate Problem \eqref{eq:transformed-5 problem formulation} as a mixed-integer semidefinite program (MISDP), which is formulatde as
{\allowdisplaybreaks 
\begin{subequations}\label{eq:adversarial problem}
    \begin{align}
 &\min_{\mathbf{G}_{m,n},\mathbf{g}_{m,n},\mathbf{u}}\tilde{F}_{ST}\left(\alpha_{m,n}, \mathbf{G}_{m,n}, \mathbf{V}, \mathbf{W}\right) \label{eq:adversarial problem-a}\\
        \mathrm{s.t.} \quad & \mathbf{T}_{m,n}^l \mathbf{g}_{m,n}^l \geq \sum_{i=1}^{Z_l} \mathbf{u}^{i,l} \underline{\tau}^{i,l},  \forall l \in [L], \label{eq:adversarial problem-b} \\
        & \mathbf{T}_{m,n}^l \mathbf{g}_{m,n}^l \leq \sum_{i=1}^{Z_l} \mathbf{u}^{i,l} \overline{\tau}^{i,l},\forall l \in [L], \label{eq:adversarial problem-c} \\
        &\mathbf{g}_{m,n}^{l+1} = \mathrm{diag}\left( \sum_{i=1}^{Z_l} \mathbf{u}^{i,l} a_i^l \right) \mathbf{T}_{m,n}^l \mathbf{g}_{m,n}^l\nonumber\\
        & + \sum_{i=1}^{Z_l} \mathbf{u}^{i,l} b_i^l , \forall l \in [L],\label{eq:adversarial problem-d}\\
        & \sum_{i=1}^{Z_l} \mathbf{u}^{i,l} = \mathbf{1}_{d_l \times 1} , \forall l \in [L],\label{eq:adversarial problem-e}\\
        &\mathbf{u}^{i,l} \in \{0,1\}^{d_l},\forall i \in [Z_l], \forall l \in [L], \label{eq:adversarial problem-f} \\
        & \text{Tr}(\tilde{\mathbf{T}}_{\mathbf{u}}^H \tilde{\mathbf{T}}_{\mathbf{u}} \mathbf{G}_{m,n}) + 2\text{Re}\{(\tilde{\mathbf{b}}_{\mathbf{u}} - \overline{c})^H \tilde{\mathbf{T}}_{\mathbf{u}} \mathbf{g}_{m,n}\} \nonumber\\ 
        &+\|\tilde{\mathbf{b}}_{\mathbf{u}} - \overline{c}\|^2 \leq R^2, \label{eq:adversarial problem-g}\\
        &\begin{bmatrix} \mathbf{G}_{m,n} & \mathbf{g}_{m,n} \\ \mathbf{g}_{m,n}^H & 1 \end{bmatrix} \succeq 0, \label{eq:adversarial problem-h}
    \end{align}
\end{subequations}}
\!\!\!where $\mathbf{g}_{m,n}^l$ represents the output of layer $l$, and constraints \eqref{eq:adversarial problem-b}-\eqref{eq:adversarial problem-f} ensure the consistency of the activation pattern.

Although the aforementioned transformed problem provides a safe robust guarantee, solving it directly is computationally demanding. The presence of binary variables $\mathbf{u}^{i,l}$ renders the ASP NP-hard. To address this computational bottleneck, we propose an efficient method by exploiting the information in the training data.

The fundamental premise of the piecewise affine neural network is that it partitions the input space into disjoint polyhedral regions, each characterized by a unique activation pattern $\mathbf{u}\in\mathcal{B}$. The uncertainty set $\mathcal{U}$ is consequently the union of these specific regions intersected with the norm constraint. Since the DNN is trained to capture the distribution of historical channel errors, the physically meaningful regions of the uncertainty set are those containing the training samples. Therefore, instead of searching through all theoretically possible binary combinations of $\mathbf{u}$, we restrict our search to the specific activation patterns that are active for the training data. 
For the historical training samples $\mathcal{D}$, we perform the following steps: First, we conduct a forward pass on the trained network for each sample in $\mathcal{D}$ and record the activation status of every neuron to determine its specific activation pattern $\mathbf{u}$. We then aggregate these patterns into a set $\mathcal{S}$, removing duplicates to obtain distinct activation patterns
\begin{equation}
    \mathcal{S}_{m,n} = \{\hat{\mathbf{u}}_{m,n}(\boldsymbol{\varepsilon}^{m,n}) \mid \boldsymbol{\varepsilon}^{m,n} \in \mathcal{D}_{m,n}\},
\end{equation}
where $\hat{\mathbf{u}}_{m,n}(\boldsymbol{\varepsilon}^{m,n})$ is the activation pattern determined by passing sample $\boldsymbol{\varepsilon}^{m,n}$ through the trained network. Consider a specific candidate pattern $\hat{\mathbf{u}}_{m,n}\in \mathcal{S}_{m,n}$, where its components $\hat{\mathbf{u}}^{i,l}_{m,n} \in \left\{0,1\right\}^{d_l}$ are now fixed binary constants indicating the active intervals for layer $l$. This definition ensures that $\mathcal{S}$ contains no redundant elements; if multiple training samples fall within the same linear region of the neural network (i.e., they generate the exact same binary vector), that pattern is included in $\mathcal{S}$ only once. This step significantly improves computational efficiency by avoiding repetitive calculations for the same polyhedral region.

Since the number of training samples $D$ is finite, the size of $\mathcal{S}$ is much smaller than the total number of possible combinatorial patterns. For a fixed activation pattern $\hat{\mathbf{u}} \in \mathcal{S}$, the binary variables in Problem \eqref{eq:adversarial problem} become fixed constants. Consequently, the constraints \eqref{eq:adversarial problem-b}-\eqref{eq:adversarial problem-f} reduce to a set of linear inequalities and equalities.
Specifically, for a fixed $\hat{\mathbf{u}}$, the ASP \eqref{eq:adversarial problem} simplifies to a convex optimization problem:
{\allowdisplaybreaks
\begin{subequations}
    \begin{align} &\min_{\mathbf{G}_{m,n},\mathbf{g}_{m,n}}\tilde{F}_{ST}\left(\alpha_{m,n}, \mathbf{G}_{m,n}, \mathbf{V}, \mathbf{W}\right)\label{eq:fixed_pattern_problem-a}\\
        \mathrm{s.t.} \quad & \mathbf{T}_{m,n}^l \mathbf{g}_{m,n}^l \geq \sum_{i=1}^{Z_l} \hat{\mathbf{u}}^{i,l}_{m,n} \underline{\tau}^{i,l},\forall l \in [L],\forall \hat{\mathbf{u}}_{m,n} \in \mathcal{S}_{m,n}, \label{eq:fixed_pattern_problem-b}\\
        &\mathbf{T}_{m,n}^l \mathbf{g}_{m,n}^l \leq \sum_{i=1}^{Z_l} \hat{\mathbf{u}}_{m,n}^{i,l} \overline{\tau}^{i,l}, \forall l \in [L],\hat{\mathbf{u}}_{m,n} \in \mathcal{S}_{m,n}, \label{eq:fixed_pattern_problem-c}\\
        &\mathbf{g}_{m,n}^{l+1} = \mathrm{diag}\left( \sum_{i=1}^{Z_l} \hat{\mathbf{u}}_{m,n}^{i,l} a_i^l \right) \mathbf{T}_{m,n}^l \mathbf{g}_{m,n}^l\nonumber\\
        &+ \sum_{i=1}^{Z_l} \hat{\mathbf{u}}_{m,n}^{i,l} b_i^l , \forall l \in [L],\label{eq:fixed_pattern_problem-d}\\
        & \text{Tr}(\tilde{\mathbf{T}}_{\hat{\mathbf{u}}_{m,n}}^H \!\tilde{\mathbf{T}}_{\hat{\mathbf{u}}_{m,n}} \!\mathbf{G}_{m,n})\! +\! 2\text{Re}\{(\tilde{\mathbf{b}}_{\hat{\mathbf{u}}_{m,n}} \!- \overline{c})^H \!\tilde{\mathbf{T}}_{\hat{\mathbf{u}}_{m,n}} \mathbf{g}_{m,n}\} \nonumber\\ 
        &+\|\tilde{\mathbf{b}}_{\hat{\mathbf{u}}_{m,n}} \!- \overline{c}\|^2 \leq R^2, \label{eq:fixed_pattern_problem-e}\\
        &\begin{bmatrix} \mathbf{G}_{m,n} & \mathbf{g}_{m,n} \\ \mathbf{g}_{m,n}^H & 1 \end{bmatrix} \succeq 0, \label{eq:fixed_pattern_problem-f}
    \end{align}
    \label{eq:fixed_pattern_problem}
\end{subequations}
}
Since Problem \eqref{eq:fixed_pattern_problem} is a convex SDP, it can be solved efficiently using standard solvers. Because the SDR inherently optimizes over an expanded PSD cone, the minimum value obtained from Problem \eqref{eq:fixed_pattern_problem} provides a global lower bound to the original non-convex ASP. The robust counterpart problem is then solved by iterating through all patterns in $\mathcal{S}$. This approach decomposes the complex MIP into finite tractable convex problems. Furthermore, since the ASP for different patterns are independent, they can be solved in parallel to further accelerate the computation.

Integrating the tractable ASP, the overall robust optimization problem is reformulated as:
{\allowdisplaybreaks
\begin{subequations}
    \label{eq:transformed-6 problem formulation} 
    \begin{align}   
        &\min_{\left\{\mathbf{V}_m\right\}\succeq0,\left\{\mathbf{W}_k\right\}\succeq0,\{\alpha_{m,n}\} }\sum_{m = 1}^{M}\mathrm{Tr}(\mathbf{V}_m)+\sum_{k = 0}^{K}\mathrm{Tr}(\mathbf{W}_k) \label{eq:t6pf-a} \\
        \mathrm{s.t.}\quad&\min_{\mathbf{G}_{m,n},\mathbf{g}_{m,n} \in \mathcal{U}_{m,n}}\tilde{F}_{ST}\left(\alpha_{m,n}, \mathbf{G}_{m,n}, \mathbf{V}, \mathbf{W}\right) \geq 0, \label{eq:t6pf-b} \\
        &\eqref{eq:t1pf-c}, \eqref{eq:t1pf-d},\eqref{eq:t1pf-g}, \label{eq:t6pf-c}
    \end{align}
\end{subequations}
}
where \eqref{eq:t6pf-b} is the ASP (\ref{eq:fixed_pattern_problem}).

After converting the chance constraints into convex constraints, i.e., the remaining challenge, the non-convex coupling between beamforming and power allocation, is addressed via an iterative framework in the next section.

\subsection{Robust Resource Management and Algorithm}
\begin{algorithm}[htbp]
\caption{Outer Loop: Precise Power Allocation Search for SAGIN}
\label{alg:outer_loop}
\begin{algorithmic}[1] 
\REQUIRE Training dataset $\mathcal{D}$, DNN uncertainty set $\mathcal{U}$, Activation patterns $\mathcal{S}_{m,n}$, Initial search grid $\mathcal{A}^{(1)}$, Refinement stages $S$, Tolerance  $\Delta\varpi^{\text{a}}$, Max iterations $I_{\max}$.
\ENSURE Optimal beamforming matrices $\mathbf{V}^*, \mathbf{W}^*$ and power coefficients $\boldsymbol{\alpha}^*$.
\STATE \textbf{Initialize:} Set stage $s = 1$.

\FOR{stage $s = 1$ to $S$}
    \IF{$s > 1$}
        \STATE Calculate new search radius $R^{(s)} \leftarrow \Delta^{(s-1)}$ (step size of previous stage).
        \STATE Generate refined grid $\mathcal{A}^{(s)}$ centered at previous best $\boldsymbol{\alpha}^{(s-1),*}$ with radius $R^{(s)}$.
    \ENDIF
    
    \FOR{each candidate $\boldsymbol{\alpha}^{(s,j)} \in \mathcal{A}^{(s)}$}
        \STATE \textbf{Execute Algorithm \ref{alg:inner_loop}:} Call the inner CPM with fixed $\boldsymbol{\alpha}^{(s,j)}$, $\Delta\varpi$, and $I_{max}$.
        \STATE \textbf{Receive:} Optimized beamforming matrices $\mathbf{V}_{s,j}$, $\mathbf{W}_{s,j}$ and current objective value $Obj_{s,j}$.
    \ENDFOR
    
    \STATE Update best candidate: $\boldsymbol{\alpha}^{*(s)} \leftarrow \argmin_{\boldsymbol{\alpha}^{(s,j)}} \{ Obj_{s,j} \}$.
    \STATE Store the corresponding optimal matrices: $\mathbf{V}^* \leftarrow \mathbf{V}_{s,j^*}$, $\mathbf{W}^* \leftarrow \mathbf{W}_{s,j^*}$.
\ENDFOR
\RETURN $\mathbf{V}^*, \mathbf{W}^*, \boldsymbol{\alpha}^*$.
\end{algorithmic}
{\footnotesize $^{\text{a}}$The parameter is set to $\Delta\varpi = -10^{-5}$ in this paper.}
\end{algorithm}

\begin{algorithm}[htbp]
\caption{Inner Loop: CPM for Robust Beamforming}
\label{alg:inner_loop}
\begin{algorithmic}[1] 
\REQUIRE Fixed power allocation coefficients $\boldsymbol{\alpha}$, Activation patterns $\mathcal{S}_{m,n}$, Tolerance $\Delta\varpi$, Max iterations $I_{max}$.
\ENSURE Optimized beamforming matrices $\mathbf{V}, \mathbf{W}$ and current objective value $Obj$.
\STATE \textbf{Initialize:} Constraint set $\mathcal{G} \leftarrow \{(\bar{\mathbf{G}}_{m,n}, \bar{\mathbf{g}}_{m,n})\}$, $i \leftarrow 0$, $Converged \leftarrow \text{FALSE}$.

\WHILE{not $Converged$ and $i < I_{max}$}
    \STATE $\mathcal{G}_{worst} \leftarrow \emptyset$
    \STATE \textbf{Step 1: Solve MP} \eqref{eq:transformed-6 problem formulation} with fixed $\boldsymbol{\alpha}$ and constraints $\mathcal{G}$.
    \STATE Obtain optimal $\mathbf{V}^{(i)}, \mathbf{W}^{(i)}$ and record the objective value $Obj$.
    
    \STATE \textbf{Step 2: Adversarial Search} via Uncertainty Set and fixed $\mathbf{V}^{(i)}, \mathbf{W}^{(i)}$:
    \FOR{each QoS constraint $(m,n)$}
        \FOR{each pattern $\hat{\mathbf{u}} \in \mathcal{S}_{m,n}$}
        \STATE Solve ASP \eqref{eq:fixed_pattern_problem} to evaluate least favorable violations.
        \ENDFOR
        \STATE Find least favorable violation $val_{m,n}$ and variables ($\mathbf{G}^*_{m,n}, \mathbf{g}^*_{m,n}$) across all patterns:
        \STATE $ val_{m,n} = \min_{\mathbf{g}} \tilde{F}_{ST}\left(\alpha_{m,n}, \mathbf{G}, \mathbf{V}^{(i)}, \mathbf{W}^{(i)}\right)$
        \STATE $(\mathbf{G}^*_{m,n}, \mathbf{g}^*_{m,n}) \!=\! \argmin_{\mathbf{g}} \!\tilde{F}_{ST}\left(\alpha_{m,n}, \mathbf{G}, \mathbf{V}^{(i)}, \mathbf{W}^{(i)}\right)$
        
        \IF{$val_{m,n} < \Delta\varpi$}
            \STATE $\mathcal{G}_{worst} \leftarrow \mathcal{G}_{worst} \cup \{ ( \mathbf{G}^*_{m,n},\mathbf{g}^*_{m,n}) \}$.
        \ENDIF
    \ENDFOR
    
    \STATE \textbf{Step 3: Convergence Check.}
    \IF{$\mathcal{G}_{worst} \neq \emptyset$}
        \STATE Update constraints: $\mathcal{G} \leftarrow \mathcal{G} \cup \mathcal{G}_{worst}$.
        \STATE $i \leftarrow i + 1$.
    \ELSE
        \STATE $Converged \leftarrow \text{TRUE}$.
    \ENDIF
\ENDWHILE
\RETURN $\mathbf{V}^{(i)}, \mathbf{W}^{(i)}, Obj$.
\end{algorithmic}
\end{algorithm}

Although the non-convexity introduced by the neural network has been resolved, Problem \eqref{eq:transformed-6 problem formulation} remains computationally intractable due to the coupling between the beamforming matrices and the power allocation coefficients $\alpha_{m,n}$. To address this, we propose a two-layer iterative algorithm that decomposes the original problem. This framework consists of an outer loop for optimizing the power allocation coefficients and an inner loop that employs the CPM to solve the robust beamforming problem. 

The outer loop focuses on optimizing the NOMA power allocation coefficients $\boldsymbol{\alpha} = \{\alpha_{m,n} \in [0, 1]| \forall m, n\}$. Since the parameters $\alpha_{m,n}$ appear in the interference terms, they are coupled with the beamforming variables. We decouple them using an alternating optimization strategy, where the power parameters and beamforming matrices are optimized sequentially. To efficiently identify the optimal power allocation coefficients $\boldsymbol{\alpha}$ without incurring the prohibitive computational cost of a dense global search, we construct a dynamic search grid, denoted as $\mathcal{A}^{(s)}$, which evolves over refinement stages $s = 0, 1, \cdots, S$. The grid construction follows a ``coarse-to-fine" strategy. Let $\alpha^{(s-1),*}$ denote the optimal parameter identified in the previous stage, where the superscript  $*$ denotes the optimal value. (For $s=0$, we set an initial center). The search grid at stage $s$ is defined as a uniform discretization of the local interval centered at $\alpha^{(s-1),*}$:
\begin{equation}
\mathcal{A}^{(s)} \!=\! \left\{\! \alpha\! \left| \alpha\! =\! \alpha^{(s-1),*}\! +\! k \Delta^{(s)}\right.\!, k \in \{- \frac{K_{grid}}{2}, \dots, \frac{K_{grid}}{2}\} \!\!\right\},
\end{equation}
where $K_{grid} = \{ 2n \mid n \in \mathbb{Z}^+ \}$, $K_{grid}+1$ is the number of grid points and the search radius, $R^{(s)}$ is adaptively updated based on the resolution of the previous stage, with the search range defined as $[\alpha^{(s-1),*} - R^{(s)}, \alpha^{(s-1),*} + R^{(s)}]$. Specifically, we set $R^{(s)} = \Delta^{(s-1)}$, where $\Delta^{(s-1)}= \frac{2R^{(s-1)}}{K_{grid}}$ denotes the step size of the previous grid. This ensures that the new search space precisely covers the gap between the current optimal solution and its nearest neighbors from the coarse grid. The power allocation search strategy is summarized in \textbf{Algorithm 1}.

As shown in \textbf{Algorithm 2}, the inner loop solves the robust beamforming problem for a specific ${\alpha}^{(s,j)}$ passed from the outer loop using the CPM, where $(s,j)$ denotes the $j$-th candidate value in the $s$-th stage. The core of this loop is the separation of the master problem (MP) and the ASP. First, for the MP, we solve a relaxed semidefinite program (SDP) considering only a finite subset of channel scenarios $\mathcal{G}$. 
Before proceeding to the ASP, it is crucial to establish the structural properties of the MP's relaxed solution. Specifically, at each iteration of Algorithm~2, the MP is Problem~\eqref{eq:transformed-6 problem formulation} restricted to the current finite cut set $\mathcal{G}$ with the power coefficients $\alpha$ fixed. Denoting the accumulated cuts for ST $(m,n)$ by the PSD data matrices $\{\bar{\mathbf{G}}_{m,n}^{(c)}\succeq\mathbf{0}\}_{c\in \mathcal{G}_{m,n}}$, the MP is a linear SDP in $\{\mathbf{V}_m\succeq \mathbf{0}\}$ and $\{\mathbf{W}_k\succeq\mathbf{0}\}$. Crucially, the DNN, the activation patterns, and the ASP relaxation enter the MP only through these fixed matrices $\bar{\mathbf{G}}_{m,n}^{(c)}$; hence the following analysis is independent of how the uncertainty set is learned.

\textbf{Assumption~1:} { \itshape  The MP is strictly
feasible. This is the standard Slater condition; It holds whenever the
QoS targets are jointly achievable, which is maintained along the
iterations since $\mathcal{G}$ only grows and the outer loop discards
any $\alpha$ with $\tilde{\alpha}_{m,n}\le 0$.}

Under Assumption~1, strong duality holds and the KKT conditions are necessary and sufficient. Let $\lambda_{m,n,c}\ge 0$, $\mu^L_k\ge 0$, $\nu^L_k\ge 0$ be the multipliers of the cut, multicast, and unicast constraints, and let $\mathbf{Z}^{\mathbf{V}}_{m}\succeq\mathbf{0}$, $\mathbf{Z}^{\mathbf{W}}_{k}\succeq\mathbf{0}$ be the PSD multipliers associated with $\mathbf{V}_m\succeq\mathbf{0}$ and $\mathbf{W}_k\succeq\mathbf{0}$. Stationarity of the Lagrangian yields the dual-slack matrices
\begin{equation}
\mathbf{Z}^{\mathbf{V}}_{m}=\mathbf{A}_m-\mathbf{S}_m,
\mathbf{Z}^{\mathbf{W}}_{0}=\mathbf{A}_0-\mathbf{S}_0,
\mathbf{Z}^{\mathbf{W}}_{k}=\mathbf{A}_k-\nu_k\mathbf{H}_k~(k\ge 1),
\end{equation}
where, using $\tilde{\alpha}_{m,n}>0$ under feasibility,
\begin{align}
\mathbf{A}_m &=\mathbf{I}+\!\sum_{i\ne m}\sum_{n,c}\lambda_{i,n,c}
\Gamma_{i,n}\bar{\mathbf{G}}^{(c)}_{i,n}\!\succ\!\mathbf{0},\notag\\
\mathbf{S}_m &=\!\sum_{n,c}\lambda_{m,n,c}\tilde{\alpha}_{m,n}
\bar{\mathbf{G}}^{(c)}_{m,n}\!\succeq\!\mathbf{0},\notag\\
\mathbf{A}_0 &=\mathbf{I}+\!\sum_{m,n,c}\lambda_{m,n,c}\Gamma_{m,n}
\mathbf{H}_{m,n}\!\succ\!\mathbf{0},~
\mathbf{S}_0=\!\sum_{k=1}^{K}\mu^L_k\mathbf{H}_k\!\succeq\!\mathbf{0},\notag\\
\mathbf{A}_k &=\mathbf{I}+\!\sum_{m,n,c}\lambda_{m,n,c}\Gamma_{m,n}
\mathbf{H}_{m,n}+\!\sum_{j=1}^{K}\mu^L_j\Gamma^{mc}_j\mathbf{H}_j \notag \\
&\quad+\sum_{j\ne k}\nu^L_j\Gamma^{uc}_j\mathbf{H}_j\!\succ\!\mathbf{0}.\notag
\end{align}
Each $\mathbf{A}_{\bullet}\succ\mathbf{0}$ since it is the identity
plus a nonnegative combination of PSD matrices. The complementary
slackness conditions $\mathbf{Z}^{\mathbf{V}}_{m}\mathbf{V}_m=\mathbf{0}$ and
$\mathbf{Z}^{\mathbf{W}}_{k}\mathbf{W}_k=\mathbf{0}$ give
$\mathrm{range}(\mathbf{V}_m)\subseteq\ker(\mathbf{Z}^{\mathbf{V}}_{m})$ and
$\mathrm{range}(\mathbf{W}_k)\subseteq\ker(\mathbf{Z}^{\mathbf{W}}_{k})$.

\textbf{Lemma~1:} { \itshape Let $\mathbf{A}\succ\mathbf{0}$,
$\mathbf{S}\succeq\mathbf{0}$, and $\mathbf{Z}=\mathbf{A}-\mathbf{S}
\succeq\mathbf{0}$. Then $\dim\ker(\mathbf{Z})\le\mathrm{rank}
(\mathbf{S})$.}

\textbf{Proof:} If $\mathbf{x}\in\ker(\mathbf{Z})\cap\ker(\mathbf{S})$,
then $\mathbf{A}\mathbf{x}=(\mathbf{Z}+\mathbf{S})\mathbf{x}=\mathbf{0}$,
so $\mathbf{x}=\mathbf{0}$ because $\mathbf{A}\succ\mathbf{0}$. Hence
$\mathbf{S}$ restricted to $\ker(\mathbf{Z})$ is injective, which gives
$\dim\ker(\mathbf{Z})\le\mathrm{rank}(\mathbf{S})$. $\blacksquare$ 

\textbf{Theorem~1 (Unicast beamformers are rank-one):} { \itshape Under
Assumption~1, every optimal solution of the MP satisfies
$\mathrm{rank}(\mathbf{W}_k)=1$ for all $k\ge 1$, irrespective of the
learned uncertainty set.}

\textbf{Proof:} Suppose $\nu_k=0$. Then $\mathbf{Z}^{\mathbf{W}}_{k}=\mathbf{A}_k
\succ\mathbf{0}$, and $\mathbf{Z}^{\mathbf{W}}_{k}\mathbf{W}_k=\mathbf{0}$ forces
$\mathbf{W}_k=\mathbf{0}$. The unicast constraint then reduces to
$-\Gamma^{uc}_k\!\left(\sum_{i\ne k}\mathrm{Tr}(\mathbf{H}_k
\mathbf{W}_i)+\sigma^2_k\right)\ge 0$, which is impossible since
$\Gamma^{uc}_k>0$ and $\sigma^2_k>0$. Hence $\nu_k>0$. Because the
subtracted term $\nu_k\mathbf{H}_k=\nu_k\mathbf{h}_k\mathbf{h}_k^{H}$
has rank-one, Lemma~1 gives $\dim\ker(\mathbf{Z}^{\mathbf{W}}_{k})\le 1$, so
$\mathrm{rank}(\mathbf{W}_k)\le 1$. Feasibility also forces
$\mathbf{W}_k\ne\mathbf{0}$, hence $\mathrm{rank}(\mathbf{W}_k)=1$. $\blacksquare$ 

\textbf{Proposition~1 (Satellite and multicast beamformers):} { \itshape
Consider the generalized eigenproblem $\mathbf{S}_m\mathbf{u}=\theta\,
\mathbf{A}_m\mathbf{u}$ (resp. $\mathbf{S}_0\mathbf{u}=\theta\,
\mathbf{A}_0\mathbf{u}$). Under Assumption~1, its largest generalized
eigenvalue equals one. If this eigenvalue is simple---equivalently,
if $\mathrm{rank}(\mathbf{Z}^{\mathbf{V}}_{m})=N_s-1$ (resp.
$\mathrm{rank}(\mathbf{Z}^{\mathbf{W}}_{0})=N_u-1$)---then the corresponding
optimal $\mathbf{V}_m$ (resp. $\mathbf{W}_0$) is rank-one, and its
principal component is the associated generalized eigenvector. This
sufficient condition is directly verifiable from the dual-slack matrix
at the optimum.}

\textbf{Proof:} From $\mathbf{Z}^{V}_{m}=\mathbf{A}_m-\mathbf{S}_m
\succeq\mathbf{0}$ with $\mathbf{A}_m\succ\mathbf{0}$, congruence by
$\mathbf{A}_m^{-1/2}$ gives $\mathbf{I}-\mathbf{A}_m^{-1/2}\mathbf{S}_m
\mathbf{A}_m^{-1/2}\succeq\mathbf{0}$. Thus the eigenvalues of
$\mathbf{A}_m^{-1/2}\mathbf{S}_m\mathbf{A}_m^{-1/2}$---equivalently the
generalized eigenvalues $\theta$ of $(\mathbf{S}_m,\mathbf{A}_m)$---are
at most one, and $\dim\ker(\mathbf{Z}^{V}_{m})$ equals the multiplicity
of $\theta=1$. Feasibility forces $\mathbf{V}_m\ne\mathbf{0}$, so
$\ker(\mathbf{Z}^{V}_{m})\ne\{\mathbf{0}\}$ and $\theta_{\max}=1$ is
attained. If $\theta=1$ is simple, then $\dim\ker(\mathbf{Z}^{V}_{m})=1$,
and $\mathrm{range}(\mathbf{V}_m)\subseteq\ker(\mathbf{Z}^{V}_{m})$
yields $\mathrm{rank}(\mathbf{V}_m)=1$; writing $\mathbf{V}_m=p\,
\mathbf{u}\mathbf{u}^{H}$ with $\mathbf{Z}^{V}_{m}\mathbf{u}=\mathbf{0}$
identifies $\mathbf{u}$ as the principal generalized eigenvector. The
argument for $\mathbf{W}_0$ is identical with $(\mathbf{S}_0,
\mathbf{A}_0)$. $\blacksquare $

When the condition in Proposition~1 holds, solving the SDR directly yields the exact rank-one optimal solution. If the condition fails and a non-rank-one solution is obtained, we recover a rank-one solution as described below.
The condition in Proposition 1 cannot be dropped for the satellite and multicast beamformers. Setting $\Gamma^{uc}_k=0$ makes $\mathbf{W}_k=\mathbf{0}$ ($k\ge 1$) optimal, and the MP reduces to $\min,\mathrm{Tr}(\mathbf{W}_0)$ s.t. $\mathbf{h}_k^{H}\mathbf{W}_0\mathbf{h}_k\ge\Gamma^{mc}_k\sigma^2_k$, $\mathbf{W}_0\succeq\mathbf{0}$, which is exactly the SDR of QoS multicast beamforming, an NP-hard problem~\cite{1634819}. An unconditional rank-one guarantee for the general model would therefore provide a polynomial-time exact solver for an NP-hard problem, which is impossible unless $\mathrm{P}=\mathrm{NP}$. Theorem 1 and Proposition 1 together constitute the strongest tightness certificate attainable for this problem: the former is unconditional for unicast, while the latter is checkable and was verified in all our experiments for the satellite and multicast beamformers. When the condition of Proposition 1 fails, we recover a rank-one solution in two stages. Deterministic rank reduction for separable SDP is applied first; it yields an optimal solution obeying the standard rank bound, far below the ambient dimension~\cite{5447068}. Gaussian randomization is used only as a final step, where the known approximation bounds of multicast SDR apply~\cite{5447068}. This hierarchy replaces the earlier heuristic reliance on randomization.

With the relaxed solution of the MP certified to be rank-one under the specified conditions, the algorithm proceeds to the ASP to identify the ``global least favorable" channel realization $\mathbf{g}^*_{m,n}$ for each user $(m,n)$ over the entire uncertainty set $\mathcal{U}_{m,n}$ to verify robustness. As derived in Section \ref{section:Formation}, $\mathcal{U}_{m,n}$ is a union of disjoint polyhedra intersected with the norm constraint, each indexed by an activation pattern $\hat{\mathbf{u}}_{m,n} \in \mathcal{S}_{m,n}$. Consequently, finding the global least favorable involves solving independent SDP for all patterns in $\mathcal{S}_{m,n}$ in parallel. The global least favorable $\mathbf{g}^*_{m,n}$ is the solution with the minimum objective value across all patterns:
\begin{equation}
    (\mathbf{G}^*_{m,n},\mathbf{g}^*_{m,n})\!=\!\argmin_{\mathbf{g} \in \mathcal{U}_{m,n}} \tilde{F}_{ST}\left(\alpha_{m,n}^{(s,j)}, \mathbf{G}_{m,n}, \mathbf{V}, \mathbf{W}\right),
\end{equation}
where $\min_{\mathbf{g} \in \mathcal{U}_{m,n}}\! \tilde{F}_{ST}\!\left(\cdot\right)\!=\!\min_{\hat{\mathbf{u}} \in \mathcal{S}_{m,n}} \{ \min_{\mathbf{g} \in \mathcal{P}(\hat{\mathbf{u}})} \!\tilde{F}_{ST}\!\left(\cdot\right) \}$. If this minimum value violates the QoS threshold (i.e., is less than tolerance $\Delta\varpi$), the corresponding $(\mathbf{G}^*_{m,n},\mathbf{g}^*_{m,n})$ is added to $\mathcal{G}$ as a new cut. This ``min-min" aggregation strategy ensures that the identified channel is indeed the most detrimental realization within the DNN-learned set, guaranteeing rigorous robustness. 
The proposed framework effectively learns CSI uncertainties and constructs asymmetric bounds to jointly optimize robust beamforming and resource allocation. The performance of the proposed method is verified through simulations in the next section.

The computational cost of the proposed framework comprises an offline DNN training stage and an online resource allocation stage. Since the offline stage is executed only once prior to deployment, the real-time efficiency is dominated by the online complexity, which involves repeatedly solving the SDP in the MP and the ASP. Let $n_{\mathrm{MP}} \triangleq M N_s + (K+1) N_u$ and $m_{\mathrm{MP}} \triangleq \vert{}\mathcal{G}\vert{} + 2K$ denote the matrix dimension and constraint count for the MP, respectively, and let $m_{\mathrm{ASP}} \triangleq \sum_{l=1}^L d_l$ denote the constraint count for the $N_s$-dimensional ASP. Based on the classical interior-point method complexity $C(n, m) = \mathcal{O}(\sqrt{n}(m n^3 + m^2 n^2 + m^3))$ for SDP \cite{5447068}, the total online complexity over $S$ search stages, $(K_{grid}+1)$ grid candidates, and $I_{max}$ cutting-plane iterations is bounded by $\mathcal{O}(S(K_{grid}+1)I_{max}[C(n_{\mathrm{MP}}, m_{\mathrm{MP}}) + (\sum_{m,n}\vert{}\mathcal{S}_{m,n}\vert{})C(N_s, m_{\mathrm{ASP}})])$.

\section{\uppercase{{\large S}imulation {\large R}esults}}
In this section, we present simulations to evaluate the performance of the proposed DNN-driven robust optimization method. A comprehensive SAGIN simulation environment is established to ensure a assessment under complex channel conditions. The key system parameters are listed in Table \ref{tab:parameters}. To validate the effectiveness of our approach, we compare it with two robust methods, including the kernel learning support vector clustering (KL-SVC) method and the ellipsoid method. Additionally, two non-robust approaches are designed to quantify the impact of channel estimation errors: the average channel gain (Arg)-based method and the worst-case channel condition (WC-CC)-based method. The Arg-based method represents a scenario of channel overestimation, where the system adopts an optimistic view of channel quality. This results in an aggressive power-saving strategy that may lead to transmission outages. Conversely, the WC-CC-based method represents a scenario of channel underestimation, where the system adopts a pessimistic view by assuming the worst possible conditions. This results in a conservative strategy that ensures connectivity but incurs excessive power consumption. These comparisons demonstrate the advantages of the proposed method in balancing reliability and efficiency.
\begin{table}[tbp]
    \centering
    \caption{System parameters}
    \label{tab:parameters}
    \begin{tabular}{|c|c|}
    \hline
    Parameter & Value \\
    \hline
    Orbit & LEO 600 km \\
    \hline
    Carrier frequency & 18 GHz \\
    \hline
    Rain fading & $\mu_\xi=-3.131$, $\sigma=1.602$ \\
    \hline
    Bandwidth & $B$=500 MHz \\
    \hline
    Noise temperature & $T$=300 K \\
    \hline
    Maximal antenna gain of ST & $G_{\text{max}}$=6 dB \\
    \hline
    Maximal beam gain & $b_{\text{max}}$=32 dB \\
    \hline
    Side-lobe level & $S_g$=-25 dB \\
    \hline
    3dB angle of satellite & $\theta_{\text{3dB}}=2.5^\circ $ \\
    \hline
    3dB angle of AP & $\varphi_x^{\text{3dB}}=60^\circ,\varphi_y^{\text{3dB}}=10^\circ$ \\
    \hline
    Antenna inter-element spacing & $d_1=d_2=\lambda/2$ \\
    \hline
    Number of NLoS paths & $L_n=5$ \\
    \hline 
    AP height & 100 m \\
    \hline
    Loss function hyperparameters & $t=5, \mu_i=5i, \nu_i=i$ \\
    \hline
    DNN architecture & 3 layers ($N \times 50$, $50 \times 50$, $50 \times 50$) \\
    \hline
    Activation function & ReLU \\
    \hline
    Learning rate & 0.00005 \\
    \hline
    LR milestone &2800 epochs \\
    \hline
    Total epochs ($I_{tr}$) & 3000 \\
    \hline
    Weight decay & $0.5 \times 10^{-8}$ \\
    \hline
    Batch size & Full dataset \\
    \hline
    \end{tabular}
\end{table}

\begin{figure}[h]
    \centering
    \includegraphics[width=0.98\columnwidth]{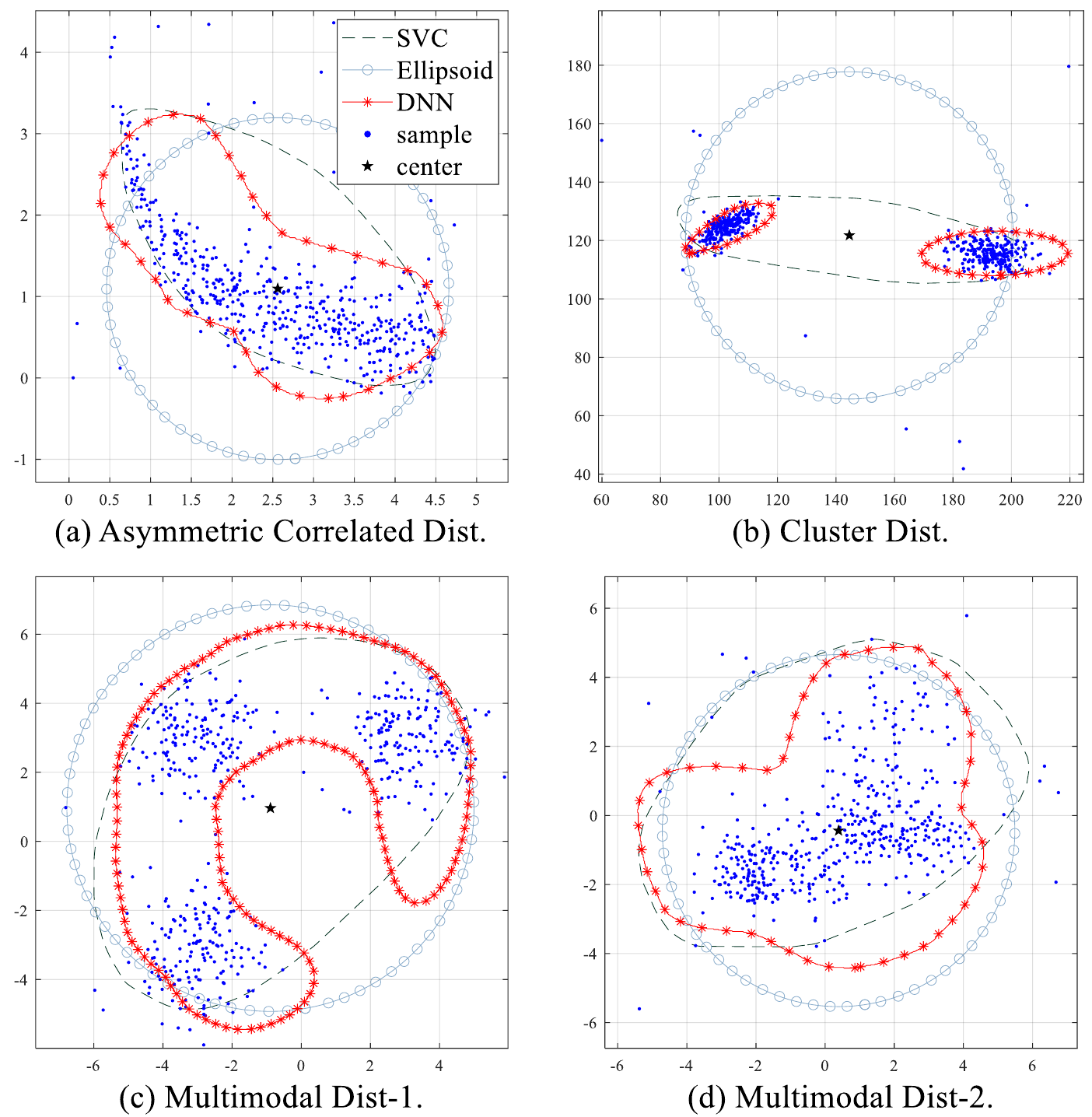}
    \caption{The uncertainty sets constructed on different datasets with $\epsilon = 0.1$ using the KL-SVC, ellipsoid-based, and DNN approaches are marked in green, azure, and red, respectively.}
    \label{fig:all_scenes} 
\end{figure}

In the first experiment, we evaluate the performance of the proposed DNN-driven uncertainty sets construction scheme under different CSI distributions. For all datasets with $\epsilon = 0.1$ using the KL-SVC, ellipsoid-based, and DNN approaches are marked in green, azure, and red, respectively. To facilitate visual analysis, the dimensions of the satellite antenna array are reduced to 2, and the resulting uncertainty sets are visualized in Fig. \ref{fig:all_scenes}.
As observed in Figs. \ref{fig:all_scenes}(a)-\ref{fig:all_scenes}(d), the geometric characteristics of the constructed sets vary significantly across methods. The ellipsoid method (indicated in azure) consistently generates a symmetric and expansive convex hull. While this ensures coverage, it fails to adapt to the irregular topologies of the underlying data, resulting in a large size of ``dead space" where no channel samples exist. The KL-SVC method (indicated in green) achieves better performance by forming a tighter boundary than the ellipsoid; however, it still exhibits limitations in capturing complex, non-convex structures. In contrast, the proposed DNN-based uncertainty set (indicated in red) demonstrates better topological adaptability. It closely wraps around the true sample distribution, accurately characterizing irregular shapes, thereby minimizing the redundant area while maintaining the required confidence level.
From a resource allocation perspective, the size of the uncertainty set is directly correlated with the conservatism of the beamforming design. A larger-sized uncertainty set means the system needs to reserve extra power to satisfy QoS constraints for worst-case scenarios that may not exist. Consequently, the ellipsoid method exhibits the highest degree of conservatism, leading to inevitable power inefficiency. The KL-SVC method reduces this conservatism to a degree but lacks the flexibility of the neural network. The proposed DNN-based approach effectively learns the hidden non-linear structure of the CSI error, significantly reducing geometric redundancy. This compactness allows the system to satisfy robustness requirements with lower power consumption, validating the effectiveness of the data-driven approach in complex channel environments.
\begin{figure}[h]
    \centering
    \includegraphics[width=1\columnwidth]{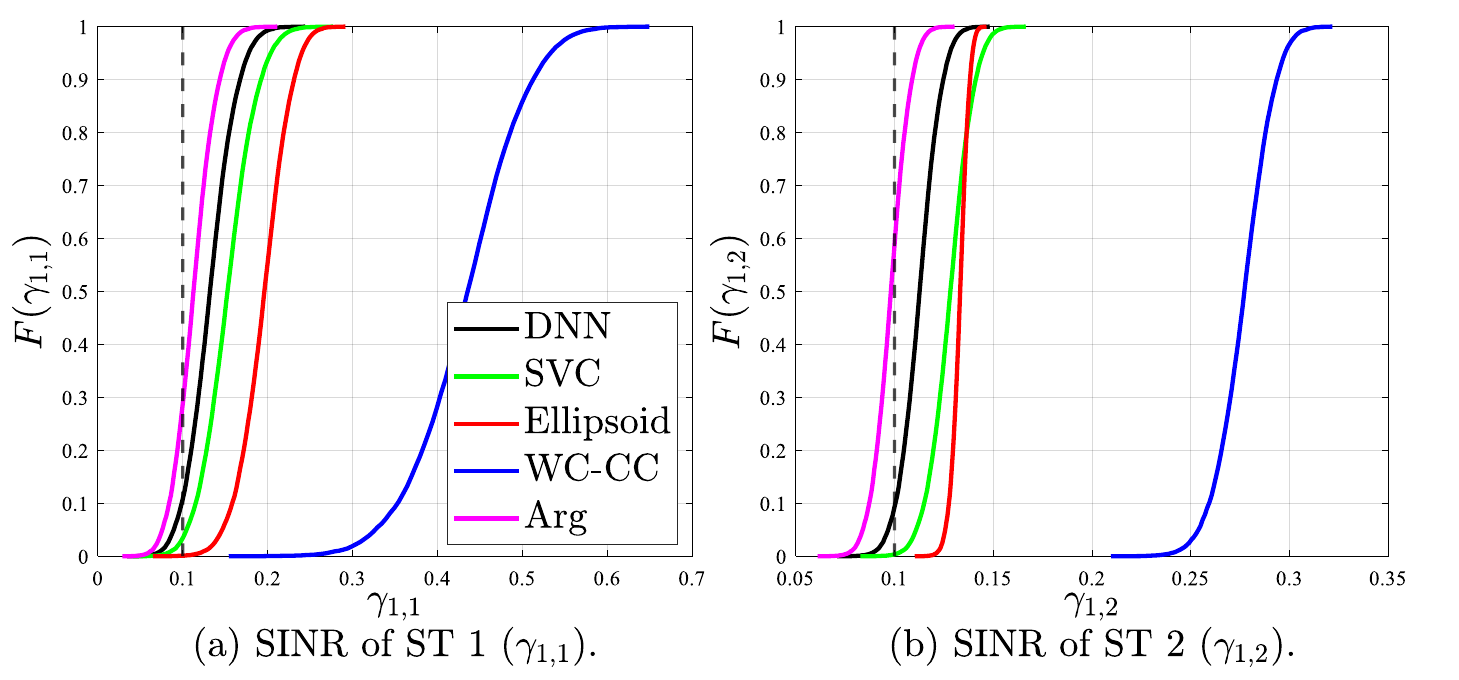}
    \caption{Cumulative distribution of the achievable SINR for STs in the first beam group, assuming $\Gamma_{1} = 0.1$, $\epsilon=0.1$.}
    \label{fig:CDF_SINR_BF1}
\end{figure}
\begin{figure}[h]
    \centering
    \includegraphics[width=1\columnwidth]{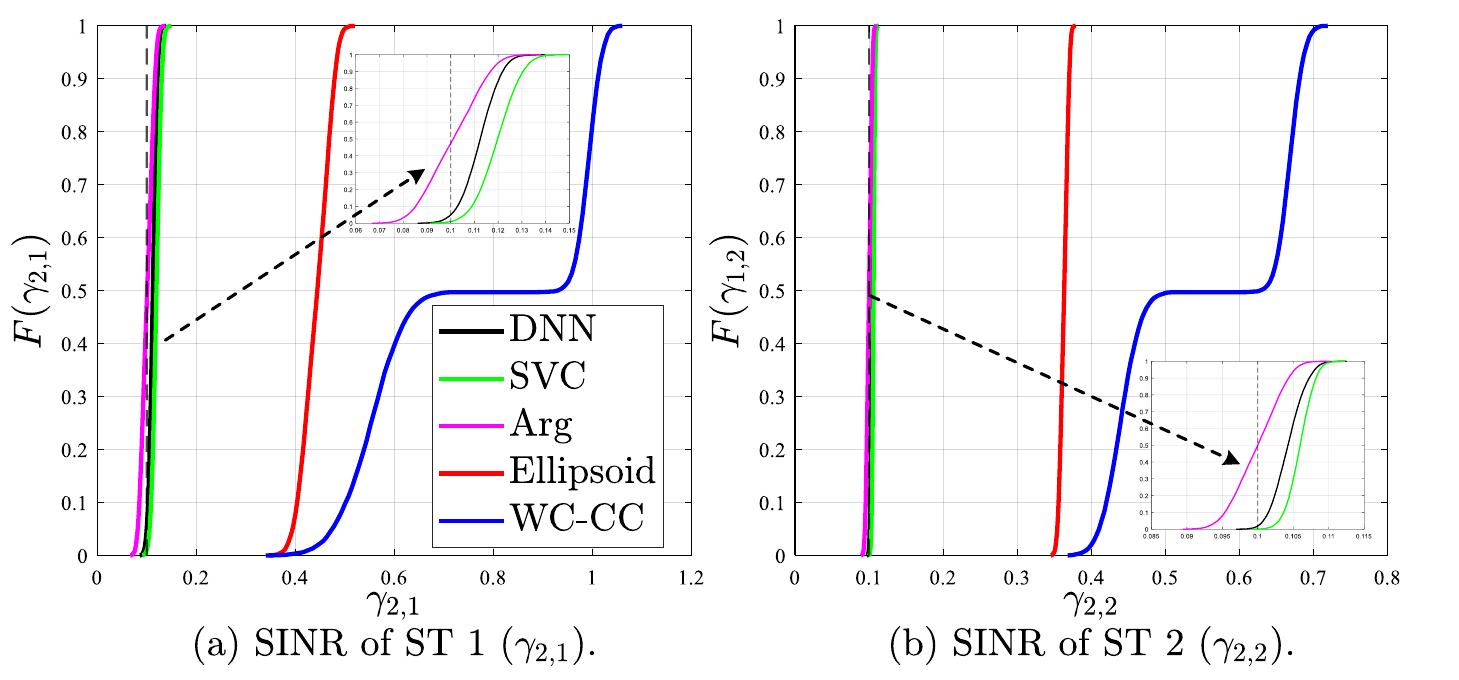}
    \caption{Cumulative distribution of the achievable SINR for STs in the second beam group, assuming $\Gamma_{2} = 0.1$, $\epsilon=0.1$.}
    \label{fig:CDF_SINR_BF2} 
\end{figure}
\begin{figure}[h]
    \centering
    \includegraphics[width=1\columnwidth]{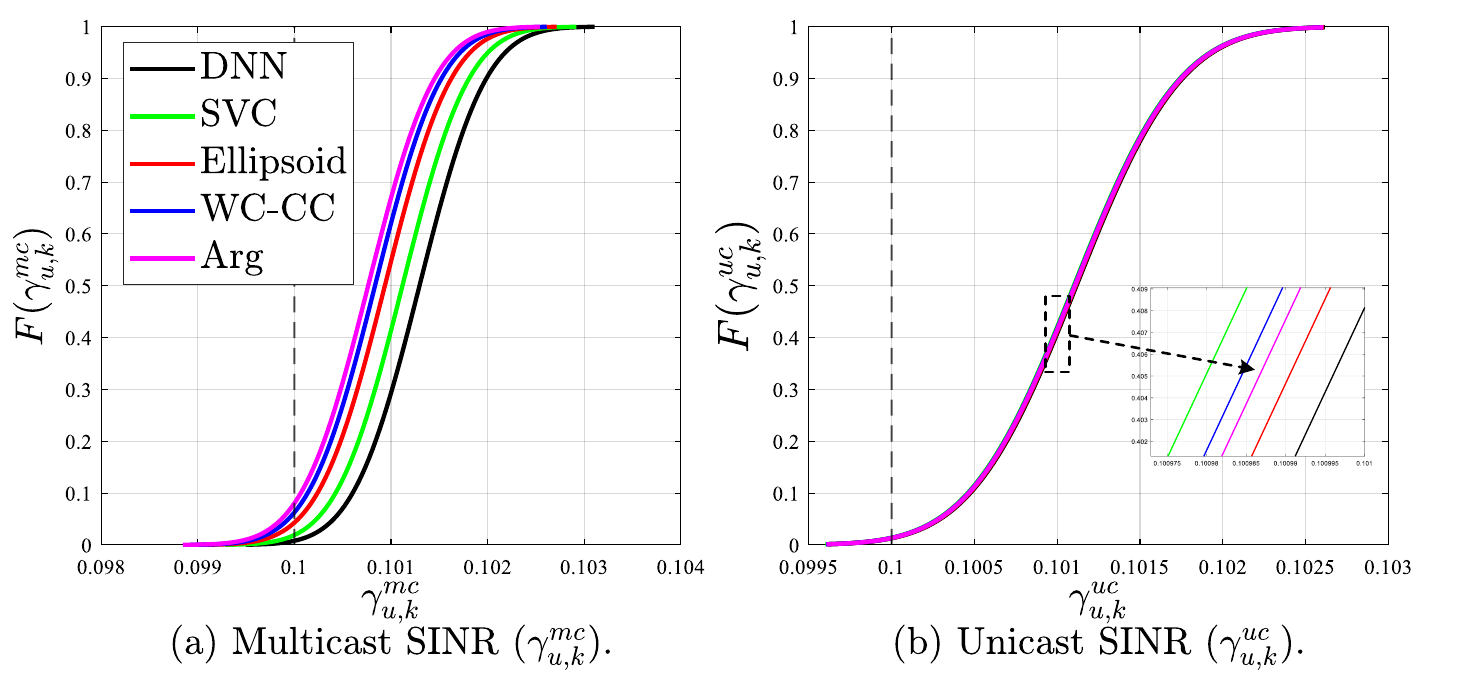}
    \caption{Cumulative distribution of the achievable SINR for AP users under perfect CSI, assuming $\Gamma_{k}^{mc}=\Gamma_{k}^{uc}=0.1$,$\epsilon=0.1$.}
    \label{fig:CDF_SINR_AP}
\end{figure}

Figs. \ref{fig:CDF_SINR_BF1} to \ref{fig:CDF_SINR_AP} present the cumulative distribution functions (CDFs) of the achievable SINR for different user groups using five distinct methods. The vertical black dashed lines in these figures indicate the predefined minimum QoS thresholds.
Fig. \ref{fig:CDF_SINR_BF1} illustrates the SINR CDF for two users within the first satellite beam group. The non-robust Arg method exhibits an overly aggressive resource allocation strategy, satisfying the SINR requirements for only about 50-60\% of the channel realizations. This inevitably leads to an unacceptable communication outage probability. On the opposite end of the spectrum, the WC-CC method adopts a strictly pessimistic approach by considering the absolute worst-case channel conditions. While it completely avoids outages, its CDF curve is shifted far to the right of the threshold, indicating an excessive over-satisfaction of QoS and a subsequent waste of transmit power. The traditional robust methods, namely the ellipsoid and KL-SVC approaches, guarantee the required outage probability but still demonstrate significant conservatism due to the geometric limitations of their symmetric structures. In contrast, the proposed DNN-based method achieves an SINR distribution that tightly hugs the preset threshold. By accurately learning and capturing the complex, asymmetric boundaries of the CSI, the DNN effectively eliminates geometric ``dead space", overcoming the conservatism of traditional methods and preventing unnecessary power consumption.
Fig. \ref{fig:CDF_SINR_BF2} depicts the SINR CDFs for two users in the second satellite beam group. A particularly noteworthy phenomenon in this scenario is the distinct step-like jump observed in the CDF curve of the WC-CC method (blue line). This unique behavior stems from the underlying CSI distribution for these specific users, which exhibits a multimodal or cluster-like topology (similar to the clustered distribution shown in Fig. \ref{fig:all_scenes}(b)). Because the WC-CC method rigidly optimizes for the single worst-case point within the uncertainty set, the presence of distinct data clusters causes the worst-case channel realization to alternate sharply between the boundaries of these separated clusters depending on the specific sample. This discontinuous optimization target directly translates into the observed sudden jumps in the received SINR. The proposed DNN method, however, constructs a highly flexible uncertainty set that precisely wraps around these separated clusters. Consequently, it mitigates the impact of extreme outliers and maintains a continuous, tight CDF curve, demonstrating superior topological adaptability in complex environments compared to both traditional robust frameworks and the worst-case baseline.
Finally, Fig. \ref{fig:CDF_SINR_AP} shows the SINR CDFs for the AP users, encompassing both multicast and unicast transmissions. Given the assumption that the AP possesses perfect CSI, the uncertainties inherent in the space-to-ground links are eliminated. As a result, the CDF curves for all five methods are virtually identical, manifesting as sharp vertical drops exactly at the preset SINR thresholds. The performance differences among the methods are negligible, indicating that in the absence of channel estimation errors, all methods can perfectly and efficiently meet the performance requirements without any redundant resource allocation.
\begin{figure}[h]
    \centering
    \includegraphics[width=1\columnwidth]{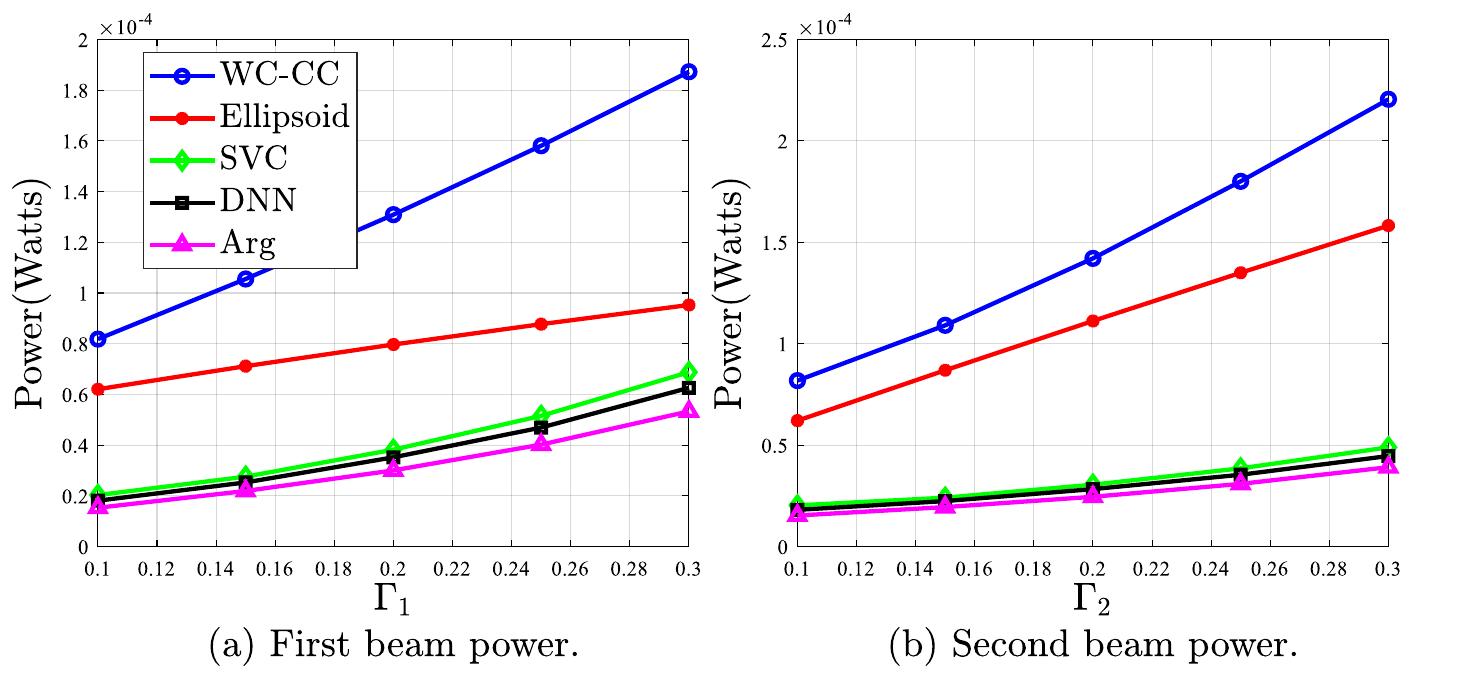}
    \caption{Total transmit power versus the predefined QoS thresholds $\Gamma$ for STs under different methods.}
    \label{fig:Total transmit power versus the predefined QoS}
\end{figure}

Fig. \ref{fig:Total transmit power versus the predefined QoS} illustrates the relationship between the total transmit power and the predefined QoS thresholds for the STs in both the first and second beam groups. As intuitively expected, a monotonic increase in total power consumption is observed across all evaluated schemes as the QoS targets become more stringent. This is because higher SINR requirements necessitate greater transmit power to overcome inherent path loss and adequately suppress the intensified cross-tier and inter-beam interference. However, the total power consumption varies significantly among the different methods. The non-robust Arg method consistently exhibits the lowest absolute power consumption. Nevertheless, as demonstrated in the previous CDF analysis, this apparent energy efficiency is deceptively achieved by adopting an overly optimistic view of the channel quality, completely ignoring estimation errors. This aggressive approach inevitably triggers severe QoS violations, rendering its low power consumption practically meaningless for highly reliable SAGIN communications. On the other extreme, the WC-CC method displays the steepest increase in power consumption as $\Gamma$ increases. By forcing the system to optimize for the absolute worst-case scenario within the uncertainty bounds, the WC-CC method suffers from severe over-provisioning of resources, leading to significant energy inefficiency.When comparing the robust optimization schemes, the traditional ellipsoid and KL-SVC methods consume considerably more power than the proposed DNN-based approach. The fundamental reason lies in their reliance on structurally symmetric geometric hulls. These rigid boundaries inevitably enclose massive non-existent ``dead space" of channel errors, forcing the resource allocation algorithm to pessimistically reserve excessive transmit power for extreme scenarios that rarely occur. In contrast, the proposed DNN method leverages the topological adaptability to tightly wrap the actual, asymmetric distribution of the CSI. By effectively eliminating geometric redundancy within the uncertainty set, the DNN-based approach avoids unnecessary power waste. Consequently, the proposed method achieves the lowest power consumption among all robust schemes, thereby demonstrating the optimal trade-off between strict communication reliability and energy efficiency.
\begin{figure}[h]
    \centering
    \includegraphics[width=1\columnwidth]{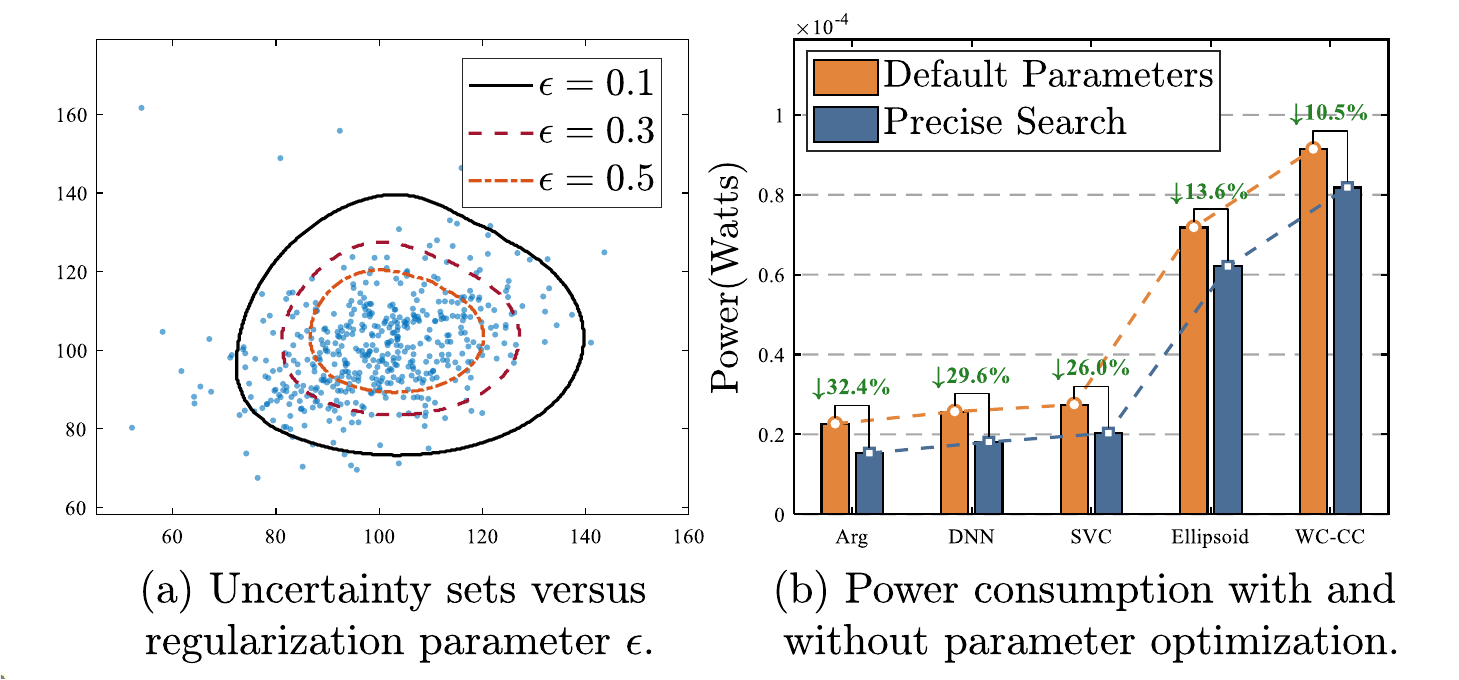}
    \caption{Performance analysis of uncertainty set adaptability and power allocation refinement.}
    \label{fig:nu_and_alpha}
\end{figure}

Fig. \ref{fig:nu_and_alpha}(a) illustrates the geometry of the learned uncertainty sets under varying maximum outage probabilities $\epsilon$. Governed by $\Pr\{\mathbf{g} \in \mathcal{U}\} \geq 1 - \epsilon$, the size of the uncertainty set exhibits a strict inverse correlation with $\epsilon$. A stringent reliability requirement (e.g., $\epsilon=0.1$) compels the DNN to expand the uncertainty boundary to encapsulate tail distributions with larger estimation errors. Conversely, relaxing the constraint ($\epsilon=0.5$) shrinks the boundary to cover only the core data distribution. This verifies the DNN's adaptability in tailoring the geometric coverage to specific QoS demands, effectively mitigating the resource over-provisioning inherent in structurally rigid uncertainty sets. Fig. \ref{fig:nu_and_alpha}(b) validates the efficacy of the proposed coarse-to-fine" search strategy. Across all evaluated schemes, dynamically optimizing the NOMA power coefficients $\alpha_{m,n}$ substantially reduces total transmit power compared to static allocations. Static assignments often mismanage the coupled relationship between $\alpha_{m,n}$ and beamforming vectors, forcing the system to compensate with higher transmit power to satisfy SINR requirements. In contrast, the precise search optimally balances intra-beam interference suppression and signal strength. Furthermore, under the optimized power allocation, the proposed DNN method achieves the lowest power consumption among all robust schemes. By eliminating geometric dead space, the DNN-generated set enables a less conservative beamforming design. While the non-robust Arg method reports lower absolute power, it incurs unacceptable QoS violations by ignoring CSI errors; conversely, the WC-CC method ensures robustness but suffers from worst-case power waste. Consequently, integrating DNN-driven uncertainty learning with precise parameter optimization achieves an optimal trade-off between energy efficiency and rigorous reliability.
\section{\uppercase{{\large C}onclusions}}
In this paper, we investigated the joint robust beamforming and resource allocation problem in SAGIN under uncertain CSI. We aimed to minimize the total transmit power while satisfying probabilistic QoS constraints. Due to the significant challenges posed by uncertain CSI in robust beamforming, a key contribution of this work was the introduction of a data-driven approach using a DNN, which constructed accurate, asymmetric  uncertainty sets to capture the complex variations in CSI. Compared with the previous methods, this approach effectively enclosed the irregular error distribution and significantly alleviated the over-conservatism in resource allocation. Furthermore, simulation results validated that the proposed method achieved an effective trade-off between energy efficiency and robustness. Additionally, we proposed a matrix property analysis method based on SDR and an efficient two-layer iterative algorithm, employing a coarse-to-fine search strategy and the CPM to convert the challenging chance-constrained non-convex problem into a solvable sequence of finite constraints.

{\small
\bibliographystyle{IEEEtran}
\bibliography{reference}
}

\end{document}